\documentclass[aps,pra,twocolumn,floatfix,longbibliography, superscriptaddress]{revtex4-1}

\usepackage{pifont}

\usepackage[dvipsnames]{xcolor}
\definecolor{bananayellow}{rgb}{1.0, 0.88, 0.21}
\definecolor{amethyst}{rgb}{0.6, 0.4, 0.8}
\definecolor{ao(english)}{rgb}{0.0, 0.5, 0.0}

\usepackage{float}
\usepackage{epsfig}
\usepackage{bm}
\usepackage[matrix,frame,arrow]{xy}
\usepackage[applemac]{inputenc}
\usepackage[T1]{fontenc}
\usepackage{lmodern}
\usepackage[english]{babel}
\usepackage{amsmath}
\usepackage{ae}
\usepackage{amssymb}
\usepackage{color}
\usepackage{graphicx}
\usepackage{url}
\usepackage{nomencl}
\usepackage{subfigure}
\usepackage{slashed}

\usepackage{booktabs}

\newcommand{\ket}[1]{|#1\rangle}

\newcommand{\secref}[1]{\mbox{Sec.~\ref{#1}}}

\renewcommand{\eqref}[1]{\mbox{Eq.~(\ref{#1})}}

\newcommand{\be}{\begin{equation}}
\newcommand{\ee}{\end{equation}}
\newcommand{\bea}{\begin{eqnarray}}
\newcommand{\eea}{\end{eqnarray}}

\usepackage{xr}
\usepackage[colorlinks]{hyperref}
\hypersetup{%
    plainpages=true,
    breaklinks=true,
    hypertexnames=false,
    pageanchor=true,
    colorlinks=true,
    linkcolor={blue},
    citecolor={red},
    urlcolor={blue},
    anchorcolor={black}
}

\newcommand{\beq}{\begin{eqnarray}}
\newcommand{\eeq}{\end{eqnarray}}

\begin{document}
	
	
	\title{Conversion of Mechanical Noise into Correlated Photon Pairs: 
	\\ Dynamical Casimir effect from an incoherent mechanical drive}
	
	
	\author{Alessio Settineri}
	\affiliation{Dipartimento di Scienze Matematiche e Informatiche, Scienze Fisiche e  Scienze della Terra,
		Universit\`{a} di Messina, I-98166 Messina, Italy}
	\affiliation{Theoretical Quantum Physics Laboratory, RIKEN Cluster for Pioneering Research, Wako-shi, Saitama 351-0198, Japan}
	\author{Vincenzo Macr\`{i}}
	\affiliation{Theoretical Quantum Physics Laboratory, RIKEN Cluster for Pioneering Research, Wako-shi, Saitama 351-0198, Japan}
	\author{Luigi Garziano}
	\affiliation{Theoretical Quantum Physics Laboratory, RIKEN Cluster for Pioneering Research, Wako-shi, Saitama 351-0198, Japan}
	\author{Omar Di Stefano}
	\affiliation{Theoretical Quantum Physics Laboratory, RIKEN Cluster for Pioneering Research, Wako-shi, Saitama 351-0198, Japan}
\author{Franco Nori}
\affiliation{Theoretical Quantum Physics Laboratory, RIKEN Cluster for Pioneering Research, Wako-shi, Saitama 351-0198, Japan}
\affiliation{Physics Department, The University of Michigan, Ann Arbor, Michigan 48109-1040, USA}
\author{Salvatore Savasta}
\affiliation{Dipartimento di Scienze Matematiche e Informatiche, Scienze Fisiche e  Scienze della Terra,
	Universit\`{a} di Messina, I-98166 Messina, Italy}
\affiliation{Theoretical Quantum Physics Laboratory, RIKEN Cluster for Pioneering Research, Wako-shi, Saitama 351-0198, Japan}
	
%
%


	\begin{abstract}
	  We show that the dynamical Casimir effect in an  optomechanical system can be achieved 
	  under incoherent mechanical pumping.
	  We adopt a fully quantum-mechanical approach for both the cavity field and the oscillating mirror. The dynamics is then evaluated using a recently-developed master equation approach in the dressed picture, including both zero and finite temperature photonic reservoirs.
	 This analysis shows that the dynamical Casimir effect can be observed even when the mean value of the mechanical displacement is zero. This opens up new possibilities for the experimental observation of this effect.
	 We also calculate cavity emission spectra both in the resonant and the dispersive regimes, providing useful information on the emission process.
	\end{abstract}
	
	\pacs{ 42.50.Pq, 42.50.Ct
	}
	
	\maketitle
	

	\section{Introduction}
 One of the most surprising predictions of  quantum field theory is that the vacuum of space is not empty, but it has plenty of short-lived {\em virtual} particles.  Real observable particles can be produced out from the quantum vacuum providing energy to its fluctuations \cite{Schwinger1951,Moore1970,Fulling1976,Yablonovitch1989,Schwinger1993}. 
Vacuum fluctuations have measurable consequences, such as the Lamb shift of atomic spectra \cite{Scully1997} and the modification of the electron magnetic moment  \cite{Greiner2008}, even when real particles are not generated. 
For years, scientists and researchers wondered if it was possible to achieve a direct observation of the virtual particles composing the quantum vacuum or, at least, if their conversion into real particles was achievable. The answer arrived only forty years ago, when Moore \cite{Moore1970} suggested that a variable length cavity undergoing relativistic motion could be able to convert virtual photons into real ones. This phenomenon was later called the dynamical Casimir effect (DCE).
Fulling and Davis \cite{Fulling1976} demonstrated that photons can be also generated by a single mirror subjected to a non-uniform acceleration. The DCE effect was first studied in the context of electromagnetic resonators with oscillating walls or containing a dielectric medium with time-modulated internal properties \cite{Dodonov1992,Dodonov1993,Ji1997,Mundarain1998}.

This concept was later generalized for other bosonic fields, e.g., cold atoms \cite{Dodonov2014}, phononic excitation of ion chains \cite{Trautmann2016}, optomechanical systems \cite{Motazedifard2017}, and Bose-Einstein condensates \cite{Carusotto2010,Jaskula2012}. Moreover, it has been  shown that photon pairs can be emitted out from the vacuum by switching or modulating the light-matter coupling strength in cavity QED systems \cite{DeLiberato2007,Anappara2009,DeLiberato2009,Garziano2013,Munoz2018}. In 1996, it was shown \cite{Lambrecht1996}  that a significant number of photons can be produced also in realistic high-$Q$ cavities with moderate mirror speeds, taking advantage of resonance-enhancement effects.  Unfortunately, the resonance conditions require the mechanical frequency $\omega_m$ to be, at least,  twice the first cavity mode frequency $\omega_c$, i.e., $\omega_m \simeq 2 n \omega_c$, where $n  \in \mathbb{N}$. This is a significant obstacle for experimental observations.

Additional theoretical studies on the DCE have been presented in, e.g., \cite{Fulling1976,Ford1982,BARTON1993,Sassaroli1994,Dodonov1996,Schaller2002,Kim2006,Dodonov2010}.
Some of these  proposals suggested to use alternative experimental setups where the boundary conditions of the electromagnetic field are modulated by an effective motion \cite{Lozovik1995,Uhlmann2004,Crocce2004,Braggio2005,Segev2007,DeLiberato2007,Souza2018}. Specifically, the link between the DCE and superconducting circuits was theoretically proposed for the first time in Ref.~\cite{Johansson2009} and elaborated later on in Ref.~\cite{Johansson2010}. In this context, the experimental results did not take long to arrive. In fact, the emission of photon pairs was observed in a coplanar transmission line terminated by a SQUID  whose inductance was modulated at high frequency \cite{Wilson2011}. The experimental realization of the DCE gives further evidence of the quantum nature of the dynamical Casimir radiation, indicating that the produced radiation can be strictly nonclassical with a measurable amount of intermode entanglement \cite{Johansson2013}. Reference \cite{Nation2012} reviews vacuum amplification phenomena with superconducting circuits.
Photon pairs were also produced by rapidly  modulating the refractive index of a Josephson metamaterial embedded in a microwave cavity \cite{Laehteenmaeki2013}. However,these do not demonstrate the conversion of mechanical energy into photon pairs, so these experiments can also be regarded as quantum simulator.

Most theoretical studies on the DCE are based on a {\it quantum} mechanical description of the electromagnetic field and a {\it classical} description of the time-dependent boundary conditions.  
Recently, the DCE in cavity optomechanical systems has been investigated without linearising the dynamics  {\it and}  describing {\it quantum}-mechanically {\it both} the cavity field and the vibrating mirror \cite{Law1995,Sala2018,macri2018}. Within this full quantum description, it turns out that the resonant generation of photons from the vacuum is determined by several ladders of mirror-field vacuum Rabi-like splittings. The resulting general resonance condition for the photon pairs production is $k\, \omega_m \simeq 2 n\, \omega_c$, ($k,n \in \mathbb{N}$). This corresponds to processes where $k$ phonons in the mechanical oscillator are converted into $ n $ cavity photon pairs. 
This generalized resonance condition enables a resonant production of photons out from the vacuum even for mechanical frequencies lower than the lowest cavity-mode frequency, thus removing one of the major obstacles for the experimental observation of this effect.

In addition, it has been shown that a vibrating mirror prepared in an excited state (mechanical Fock state) can spontaneously emit photons like a quantum emitter. In this case, however, a photon pair is emitted instead of a single photon.

Moreover, it has been recently demonstrated that virtual Casimir photon pairs can be used to enable a coherent motional coupling between two spatially separated moveable mirrors,  allowing this kind of optomechanical system to also operate as a mechanical parametric down-converter even at very weak excitations \cite{DiStefano2019a}.
Entangled photons from the vacuum can be also generated by using  microwave
circuit-acoustic resonators \cite{Wang2018}.

The approach considered in Ref.~\cite{macri2018} also extends the investigation of the DCE to the optomechanical ultrastrong-coupling (USC) regime, where the optomechanical coupling rate is comparable to the mechanical frequency \cite{Kippenberg1172,Nunnenkamp2011,Marquardt2009,Teufel2011,Chan2011,Rimberg2014,Aspelmeyer2014}. This regime, which attracted great interest also in cavity QED  giving rise to a great variety of novel quantum effects \cite{Ridolfo2012,Garziano2013,Garziano2015,Garziano2016}, turned out to be an essential feature for the realization of new interesting proposals in quantum optomechanics \cite{Garziano2015b,Macri2016,Butera2018}.

Temperature effects also play an important role for the generation of photons in a resonantly vibrating cavity \cite{PLUNIEN1987,Plunien1998,Plunien2000,MOTAZEDIFARD2018}. 
Specifically, it turns out that the thermal contributions in these systems under the influence of time-dependent boundary conditions leads to a strong enhancement of photon pairs production at finite temperatures.


Encouraged by the results obtained in Ref.~\cite{macri2018}, here we investigate the dynamics of an optomechanical system in a fully quantum-mechanical framework,
under incoherent mechanical excitation, using a master equation approach. This allows to demonstrate that a remarkable Casimir photon pairs flux is produced even considering a thermal-like noise source coupled only to the mechanical degree of freedom. 
For ultra-strongly coupled hybrid quantum systems \cite{DiStefano2017,Stassi2017,Kockum2017,Kockum2018,Forn-Diaz2018a}, the standard quantum-optical master equation breaks down, and a dressed master equation approach is needed \cite{Beaudoin2011,Ridolfo2012,Hu2015}.
Furthermore, if the energy level spectrum displays a quasi-harmonic behaviour \cite{Marquardt2009}, like in optomechanical systems, a new dressed master equation \cite{Ma2015,settineri2018} not involving the usual secular approximation is required.

The outline of this article is as follows: in \secref{II} we briefly introduce the theoretical model and the dressed master equation approach for quasi-harmonic hybrid systems. Section~\ref{sec:III} is devoted to the presentation of the energy-level structure, focusing the attention on the avoided level crossings giving rise to the DCE. In \secref{sec:IV} we apply the generalized master equation \cite{settineri2018} to calculate the dynamics of the system at finite temperatures and, using the quantum regression theorem, we present the power spectra in the weak and strong light-matter coupling regimes. We conclude in \secref{sec:VI}.

	\section{Model}
	\label{II}
	\begin{figure}
	\centering
	\includegraphics[width = 9 cm]{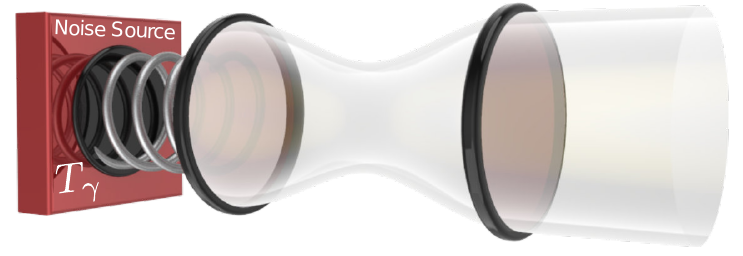}
	\caption{Schematic of a generic optomechanical system. One of the mirrors  of the optical cavity is coupled to a noise source with effective temperature $T_{\gamma}$ and can vibrate at frequency $\omega_m$. This system can generate Casimir photon pairs.
		\label{fig:1}}
\end{figure}
We study a standard optomechanical system constituted by an optical cavity with a movable end mirror [see Fig.~\ref{fig:1}]. Moreover, we consider a radiation pressure coupling between the first cavity mode and a single mechanical mode. 

The system Hamiltonian  can be written as:
	\begin{equation}\label{Hs}
	\hat H_{S} = \hat H_0 + \hat V_{\rm om} + \hat V_{\rm DCE}\, ,
	\end{equation}
where {($\hbar=1$ throughout the paper)}
	 \begin{equation}\label{H0}
	 \hat H_0 = \omega_c \hat a^\dag \hat a +  \omega_m \hat b^\dag \hat b
	 \end{equation}
is the uncoupled Hamiltonian, and 
	\begin{equation}
	\hat V_{\rm om} =g \hat a^\dag \hat a\,(\hat b + \hat b^\dag)
	\end{equation}
is the standard optomechanical interaction Hamiltonian.
Here, $\omega_c$ is the resonator frequency, $\omega_m$ is the mechanical frequency, $g$ is the optomechanical coupling strength and $\hat a$($\hat b$), $\hat a^\dag$($\hat b^\dag$) are, respectively, the bosonic creation (annihilation) operators for the cavity and mechanical modes  Finally, the perturbation term determining the DCE is
	\begin{equation}\label{VDCE}
	\hat V_{\rm DCE} = \frac{g}{2} (\hat a^2 + \hat a^{\dag 2} ) (\hat b + \hat b^\dag)\, .
	\end{equation}
Since in this case the $\hat V_{\rm DCE}$ term  only couples bare states having energy differences $2\omega_c \pm  \omega_m$ much larger then the coupling strength $g$, it can be neglected. Also, this interaction term is often neglected when describing most of the  experimental optomechanical systems, where the mechanical frequency is much smaller than the cavity frequency. 

The resulting total Hamiltonian conserves the photon number and can be diagonalized separately in each $n$-photon subspace. The general quantum state of such system is:
	\begin{equation}
	| n, k_n \rangle=| n\rangle \otimes \hat D(n\eta)| k \rangle\, ,
	\end{equation}
where the integer $ k_n $ represents the vibrational excitations of the mechanical resonator in the corresponding $ n $-photon subspace, and 
\be
| k_n \rangle= \hat{D}(n\eta)| k \rangle
\ee
represents the displaced mechanical Fock state determined by the displacement operator $\hat D(n\eta)$, where
\be
\eta\equiv g/\omega_m \,
\ee
 is the normalized coupling strength. In the manifold with $ n = 0 $, the states $ |0,k_0\rangle $, simply labeled $ |0,k\rangle $, are the eigenstates of the harmonic oscillator decoupled from the cavity.
When considering ultrahigh-frequency mechanical oscillators with resonance frequencies 
\be
\omega_m \simeq \omega_c\,,   
\ee
the $\hat V_{\rm DCE}$ term cannot be neglected. In this case, the photon number is no longer conserved and there is no analytical solution for the system eigenstates. Moreover, it turns out that the introduction of the $\hat V_{\rm DCE}$ term increases the degree of anharmonicity, slightly modifying the levels structure but still preserving the quasi-harmonic behaviour. Consequently, the system dynamics has to be described using a generalized master equation developed without performing the usual secular approximation. A suitable approach, able to describe the time evolution of the density matrix operator $\hat \rho$ for any hybrid quantum system in the presence of dissipations and thermal-like noise, has been presented in Ref.~\cite{settineri2018}. 

In the interaction picture, this master equation can be written as 
\begin{equation}\label{MET}
\dot{\hat\rho} = \kappa {\cal L}[\hat A]\hat\rho+ \gamma {\cal L}[\hat B]\hat\rho\,.
\end{equation}
with $ \kappa $ and $ \gamma $, respectively, the cavity and mirror damping rates. The dressed photon and phonon lowering operators $ \hat O= \hat A, \hat B $ are defined in terms of their corresponding bare operators  $\hat o= \hat a, \hat b $ by  the relation \cite{Ridolfo2012,Garziano2013}
\begin{equation}\label{operators}
\hat O(\omega)=\sum_{\epsilon-\epsilon^{\prime}=\omega}\hat{\Pi}(\epsilon)(\hat o+\hat o^{\dag}) \hat \Pi(\epsilon^{\prime})e^{-i\omega t}\,,
\end{equation}
where $\epsilon$ are the eigenvalues of $\hat H_S$  and $\hat \Pi (\epsilon)\equiv |\epsilon\rangle \langle \epsilon|$ indicate the projectors onto the respective eigenspaces. Furthermore, the Liouvillian superoperator ${\cal L}[\hat O]\hat\rho$ can be expressed in the general form:
\begin{equation}\label{liouOPTB} 
\begin{split}
{\cal L}[ \hat{O}]\hat \rho&=\sum_{(\omega,\omega')>0}\frac{1}{2}\bigg \{ n(\omega^{\prime},T) [\hat{O}^{\dag}(\omega^{\prime}) \hat\rho \hat{O}(\omega)-\hat{O}(\omega)\hat{O}^{\dag}(\omega^{\prime})\hat \rho  ]\\
&+[n(\omega,T)+1][\hat{O}(\omega) \hat\rho \hat{O}^{\dag}(\omega^{\prime})-\hat{O}^{\dag}(\omega^{\prime})\hat{O}(\omega)\hat\rho] \\
&+ n(\omega,T)[\hat{O}^{\dag}(\omega^{\prime}) \hat\rho \hat{O}(\omega)-\hat\rho \hat{O}(\omega)\hat{O}^{\dag}(\omega^{\prime})]\\
&+[n(\omega^{\prime},T)+1][\hat{O}(\omega) \hat\rho \hat{O}^{\dag}(\omega^{\prime})-\hat\rho \hat{O}^{\dag}(\omega^{\prime})\hat{O}(\omega)] \bigg \}\,
\end{split}
\end{equation}
where $ (k_B=1) $
\begin{equation}
n(\omega,T)=\left[{\rm exp}({\omega/T})-1\right]^{-1}\,
\end{equation}
is the thermal noise occupation number of the system reservoir, at real or {\em effective} temperature $T$.
 
When counter-rotating terms are taken into account in the interaction Hamiltonian, the introduction of master equations in the dressed basis is not sufficient. Indeed, a modification of input-output relationships, relating the intracavity field with the external fields \cite{Ridolfo2012,macri2018,settineri2018,Portolan2008,stefano2001}, is also required. According to these modified relationships, the output fields are no more determined by expectation values of the bare photon operators (see, e.g., \cite{Gardiner1985,Savasta2002,DiStefano2018}), but by the expectation values of the dressed operators in \eqref{operators}.

	\section{Vacuum Casimir-Rabi Splittings }
	\label{sec:III}
\begin{figure}
		\centering
		\includegraphics[width = 7 cm]{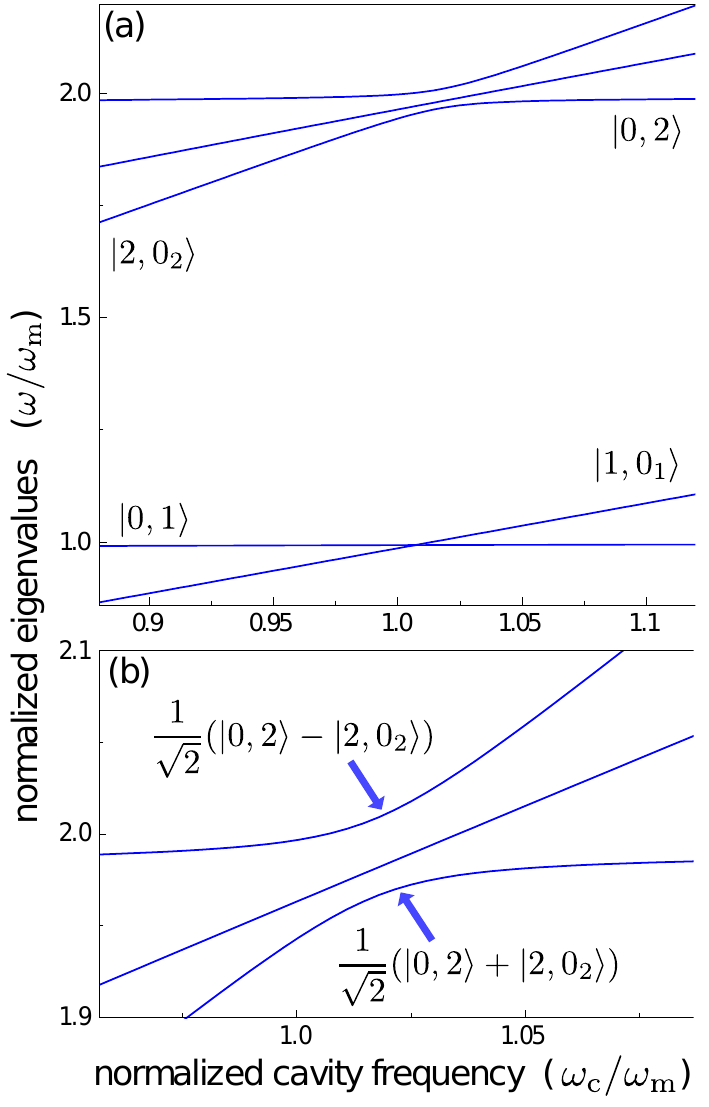}
		\caption{(a) Lowest energy eigenvalues of the system as a function of $\omega_{c}/\omega_{m}$ for a normalized optomechanical coupling strength $\eta =g/\omega_m= 0.1$. The ground state is not displayed. (b) Enlarged view of the avoided level crossing arising from the coherent coupling between the states $| 0, 2 \rangle$ and $| 2, 0_{2} \rangle$. The energy splitting reaches its minimum at the  resonant frequency $\omega_c \simeq \omega_m$.
			\label{fig:2}}
	\end{figure}
In order to fully characterize our system, we numerically diagonalize the Hamiltonian $\hat H_{S}$ in Eq.~(\ref{Hs}). Figure~\ref{fig:2}(a) shows the lowest energy levels as a function of the cavity frequency $\omega_c/\omega_m$ considering  a normalized optomechanical coupling strength $\eta= 0.1$.

As reported in Ref.~\cite{macri2018}, when the resonant conditions 
\be
q\,\omega_{m}=2 \, \omega_{c}\,
\ee
are satisfied, the $\hat V_{\rm DCE}$ term induces a coherent resonant coupling between the bare states $|0,k\rangle$ ({\it i.e.}, $0$ photons and $k$ phonons) and $|2,(k-q)_{2}\rangle$ ({\it i.e.}, $2$ photons and $k-1$ phonons), with  $q \in \mathbb{N}^*$, having different number of  excitations. Figure~\ref{fig:2}(b) shows an enlarged view of the avoided level crossing arising for $\omega_{m} \simeq  \omega_{ c}$, involving the states $| 0, 2 \rangle$  and $| 2, 0_{2} \rangle$. When the splitting is at its minimum, the two system eigenstates are essentially a symmetric and antisymmetric linear superpositions of these bare states $ | \psi_{\pm} \rangle \simeq \frac{1}{\sqrt{2}} (|0,2 \rangle \pm |2, 0_{2} \rangle )$. The size of this avoided level crossing (Casimir-Rabi splitting), analytically calculated using first-order perturbation theory, is given by
\be\label{V}
\begin{split}
2 \Omega_{0,2}^{2,0} & =2  \langle 0, 2| \hat V_{\rm DCE} | 2, 0_2 \rangle\\
&= \sqrt{2}\,   g\left[\sqrt{3}  D_{3, 0}(2 \eta) 
+  \sqrt{2} D_{1, 0}(2 \eta) \right] \, , 
\end{split}
\ee
where
\be
D_{k',k}(2 \eta)= \sqrt{k!/k'!} (2 \eta)^{k'-k} e^{-|2 \eta|^2/2} L_k^{k'-k}(|2 \eta|^2)
\ee
represents the overlap between different displaced mechanical Fock states and $L_k^{k^\prime -k}$ is an associated Laguerre polynomial. It is important to note that the quantity $2\Omega_{0,2}^{2,0}$ plays a fundamental role in the DCE, since it determines the {\it rate} at which a mechanical {\it two}-{\it phonon} state is able to {\it generate photon pairs}. Specifically, for a normalized optomechanical coupling $\eta=0.1$ we obtain a matrix element $2 \Omega_{0,2}^{2,0}\simeq0.05$ that ensures that this avoided level crossing is able to produce a detectable rate of Casimir photon pairs.
\section{RESULTS}
	\label{sec:IV}
Here, we present the system dynamics numerically evaluated taking into account a thermal-like pumping of the mechanical components and considering the photonic reservoir both at $T_{\kappa}=0$ and at finite temperature. 
Specifically, we study the time evolution of the mean phonon (photon) number $ \langle \hat B^{\dag} \hat B \rangle$ ($ \langle \hat A^{\dag}\hat A \rangle $) and the zero-delay phononic (photonic) normalized second-order correlation function, defined as
\begin{equation}
g^{(2)}_{\rm O}(t,t) = \frac{\langle \hat O^\dag(t) \hat O^\dag(t) \hat O(t) \hat O(t) \rangle}{\langle \hat O^\dag(t) \hat O(t) \rangle^2 }\, ,
\end{equation}
with $\hat{O}\in[\hat{A},\hat{B}]$.
	
\subsection{System dynamics in the weak-coupling regime}	
We start considering the system initially prepared in its ground state and in the {\em weak-coupling regime}, which corresponds to the case where the Casimir-Rabi splitting $2\Omega_{0,k}^{2,k-q}$ is smaller than the total decoherence rate  of the system $\Gamma_{\rm tot}=\gamma + \kappa$.
 \begin{figure}
	\centering
	\includegraphics[width = 8 cm]{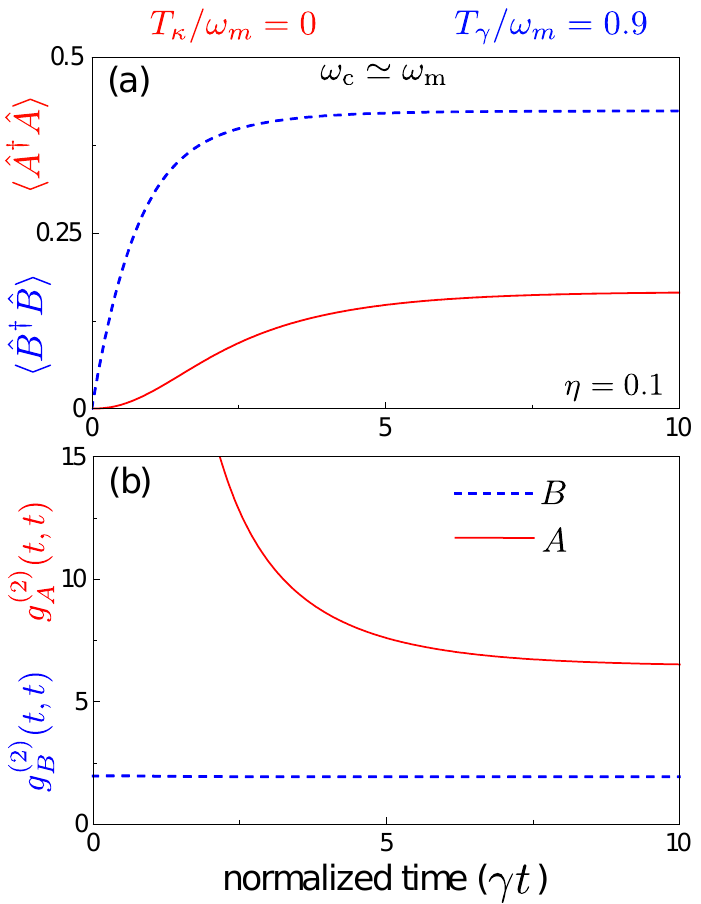}
	\caption{System dynamics for the resonant case $\omega_c \simeq  \omega_m$ considering a $T_{\kappa}=0$ cavity reservoir and the mechanical oscillator coupled to a thermal-like noise source with  an effective temperature $T_{\gamma}/\omega_{m}=0.9$. 
		(a) Time evolution of the mean phonon number $\langle \hat B^\dag \hat B \rangle$ (blue dashed curve) and of the mean intra-cavity photon number $\langle \hat A^\dag \hat A \rangle$ (red solid curve). Due to the thermal-like pumping, the populations reach  stationary values. (b) Time evolution of the zero-delay normalized photon-photon $g^{(2)}_A(t,t)$  and phonon-phonon $g_B^{(2)}(t,t)$ correlation functions. At $ t=0 $, the two-photon correlation function $g_A^{(2)}(t,t)$ displays values much higher than two, showing that a considerable number of photon pairs are emitted. As the time goes on, this value decreases significantly due to the cavity losses and the corresponding increase of the mean photon number. On the contrary, the mechanical correlation function  sets on a constant value $g_{B}^{(2)}(t, t) \approx 2$, showing that the mechanical system is in an incoherent state produced by the thermal-like noise.
} \label{fig:3}
\end{figure}

\begin{figure*}
	\centering
	\includegraphics[width =16 cm]{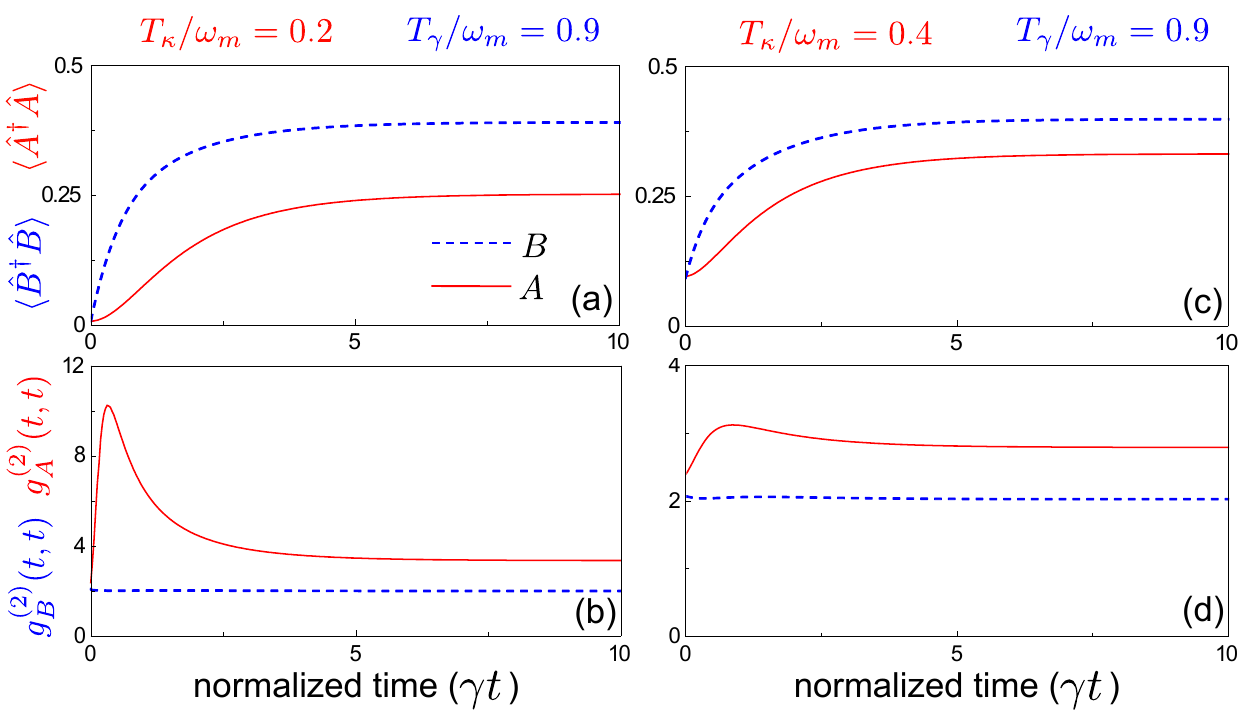}
	\caption{System dynamics evaluated considering finite-temperature reservoirs and the system initially prepared in a thermal state at the same temperature $ T_\kappa $ of the photonic reservoir. Panels (a) and (c) show the time evolution of the cavity mean photon number $\langle \hat A^\dag \hat A \rangle$ (red solid curves) and the mean phonon number $\langle \hat B^\dag \hat B \rangle$ (blue dashed curves) for  $ T_\gamma / \omega_m=0.9 $  and $ T_{\kappa}/\omega_{ m}=0.2 $ (a), $ 0.4 $ (c).
		Panels (b) and (d) display the time evolution of the zero-delay two-photon (red solid curves)  and two-phonon  (blue dashed curves)  correlation functions, $g^{(2)}_A(t,t)$ and $g^{(2)}_B(t,t)$,  for  $ T_\gamma /\omega_m=0.9  $  and $ T_{\kappa}/\omega_{ m}=0.2 $ (b), $ 0.4 $ (d).	\label{fig:4}}
\end{figure*}

Specifically,  we assume $\gamma/\omega_m=0.05$ and $\kappa=\gamma/2$ with an optomechanical coupling $\eta=0.1$, considering the resonant case $\omega_m \simeq \omega_c$ corresponding to the minimum splitting of the avoided level crossing arising between the states $| 0, 2 \rangle$  and $| 2, 0_{2} \rangle$ (see Figure~\ref{fig:2}(b)). 
Figures~\ref{fig:3}(a,b) display the time evolution of the photonic $\langle \hat A^\dag \hat A \rangle$ (red solid curve) and phononic $\langle \hat B^\dag \hat B \rangle$ (blue dashed curve) populations, together with the time evolution of the respective two-photon and two-phonon correlation functions $g_{B(A)}^{(2)}(t, t)$. All these quantities have been evaluated taking into account the interaction with a zero temperature ($T_{\kappa}=0$) photonic reservoir and providing an {\it incoherent thermal-like} pumping of the mechanical component by means of phononic reservoir with effective temperature $T_{\gamma}/\omega_{m}=0.9$. As shown in Figure~\ref{fig:3}(a), the photonic and phononic populations start from zero and, due to the incoherent thermal-like pumping of the mechanical modes, reach a considerable stationary value.
In particular, a steady state intracavity mean photon number $\langle \hat A^\dag \hat A \rangle_{\rm ss} \simeq 0.15$ is obtained. For a cavity mode of frequency $\omega_{c}/2\pi\simeq 6$ GHz, this value corresponds to a steady-state output photon flux $\Phi = \kappa \langle \hat A^\dag \hat A \rangle_{\rm ss} \sim  1.4 \times 10^8$ photons per second. 
This output photon flux is remarkable since it is much higher than the detection threshold of the state-of-the-art detectors, despite the quite low quality factor $Q_c =  \omega_c / \kappa = 40$ of the cavity considered in the numerical calculations. Furthermore, also the mechanical loss rate $\gamma$ corresponds to a quality factor $Q_m$ one order of magnitude lower than the values which are experimentally measured in ultra-high-frequency mechanical resonators \cite{OConnell2010,Rouxinol2016}. Moreover, in Fig.~\ref{fig:3}(b) we observe that the photonic correlation function starts from a value much higher than two, suggesting that a high number of photon pairs is produced. As the time goes on, this value decreases significantly due to the system losses  and  the corresponding increase of the mean photon number (note that $g_{A}^{(2)}(t, t)$ is inversely proportional to the square of the mean photon number). On the contrary, the mechanical correlation function sets on a constant value $g_{B}^{(2)}(t, t) \approx 2$, showing that the mechanical component is in an incoherent state produced by the thermal-like pumping. These results are particularly interesting since they demonstrate that the DCE can also be experimentally observed exciting a movable mirror with an {\it incoherent thermal-like pump} as, for example, a white noise generator (made by an ultra-high frequency resonator interacting with a microwave cavity).
	\begin{figure}
	\centering
	\includegraphics[width =8 cm]{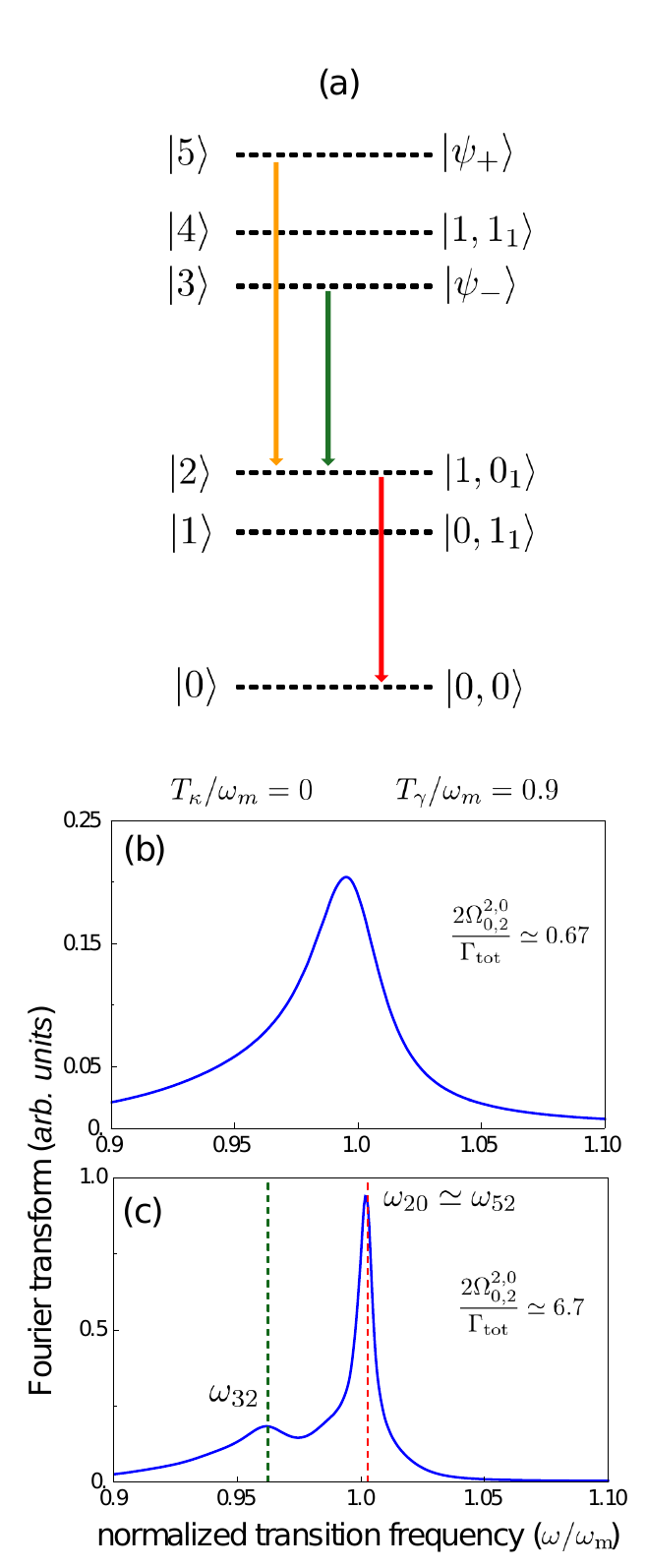}
	\caption{(a) Scheme of the first energy levels of the optomechanical system. Solid arrows represent the possible one-photon decay channels when the  effective temperature of the mechanical reservoir is high enough to populate the state $\ket{5}  $.
		Panels (b) and (c) display, respectively,  the cavity emission spectra for the system  in the weak and strong light-matter coupling regime and at zero detuning. In both cases,  the cavity reservoir is at  $ T_{\kappa}=0 $, while the mechanical oscillator is coupled to a reservoir with effective temperature $ T_\gamma/\omega_m=0.9 $.
		Parameters are: $ \omega_{ m}=1 $, $ \eta=0.1 $. The total loss rate $ \Gamma_{\rm tot}=\kappa+\gamma $ of the system is: (b)  $ 7.5 \times 10^{-2} \omega_{ m} $   and (c)  $ 7.5 \times 10^{-3} \omega_{ m} $.
		\label{fig:5}} 
\end{figure}
\begin{figure}[h!]
	\centering
	\includegraphics[width =7 cm]{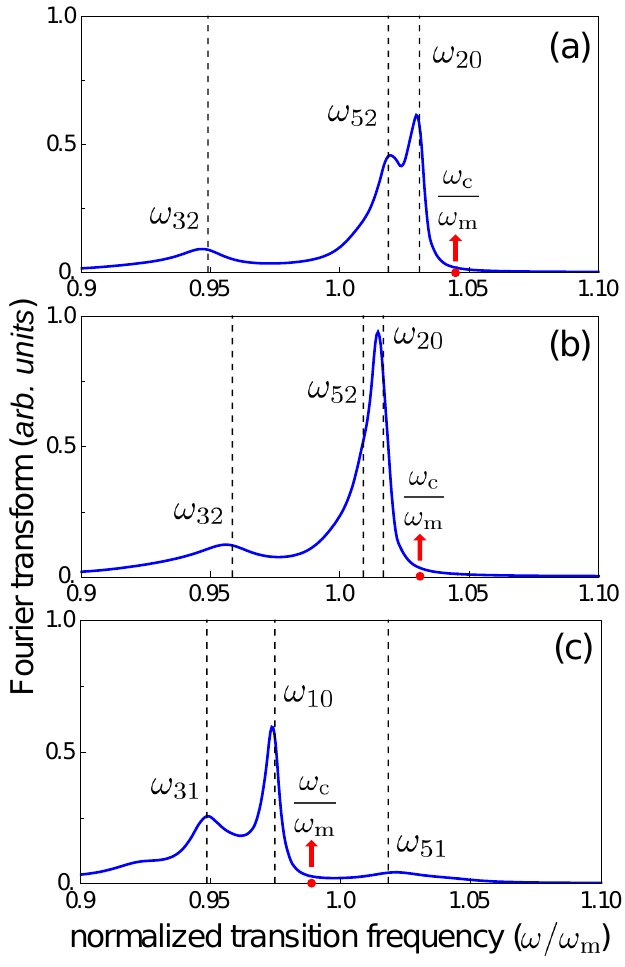}
	\caption{Cavity emission spectra for the system in the strong coupling regime for different values  of the detuning $ \Delta\equiv (\omega_{ c}-\omega_{ r})/\omega_{ m}$, where $ \omega_{ r}\simeq1.017\, \omega_{ m}$ is the frequency corresponding to the minimum value of the splitting in Fig.~\ref{fig:2}(b). Specifically, we considered the cases: (a)  $ \Delta=0.028 $, (b) $ \Delta=0.01 $ and (c) $ \Delta=-0.028 $.  Main contributions are indicated by dashed lines. Parameters are the same of Fig.~\ref{fig:5}(c). 
		\label{fig:6}}
\end{figure}
In real optomechanical systems ground state cooling is never complete and the interaction with a finite-temperature reservoir has to be taken into account. The time evolution of the photonic and phononic populations together with the respective  two-photon and two-phonon zero-delay correlation functions are displayed in Fig.~\ref{fig:4}. These functions are evaluated in more realistic conditions, taking into account non-zero temperature reservoir for both the subsystems.
In these conditions, both the populations start from a non-zero value corresponding to the initial thermal equilibrium density matrix. As expected, a fraction of the observed photons is thermal and does not originate from the mechanical to optical energy conversion mechanism.
This picture is confirmed by comparing the dynamics of the two correlation functions shown in Fig.~\ref{fig:4}(b,d). Specifically, when the cavity temperature increases, we observe a strong decrease of the $g_{A}^{(2)}(t, t)$ peak value, indicating that less photons are emitted in pairs. However, as expected, the phonon-phonon correlation functions remain constant at the thermal value $g_{B}^{(2)}(t, t) \simeq 2$.  
These results demonstrate that, when the presence of a cavity thermal noise is taken into account, the number of the produced Casimir photon pairs decrease. However, the output photon flux is still above the detection threshold of the photodetector and the peak value of the $g_{A}^{(2)}(t, t)$ indicates that photon pairs are produced.

\subsection{Emission spectra in the weak and strong coupling regimes}
In order to obtain more information on the ongoing physics, here we present the cavity emission spectra derived via a quantum regression approach. Considering a normalized optomechanical coupling $ \eta=0.1$, we present results for the system both in the weak and in the strong light-matter coupling regime for different values of $ \omega_{\rm c}/\omega_{ m} $. We consider  the cavity at  $ T_{\kappa}=0 $, while the mechanical oscillator is coupled to a reservoir with effective temperature $ T_\gamma/\omega_m=0.9 $. For the sake of simplicity, we indicate the energy eigenvalues and eigenstates as  $ \omega_{l}$ and $ \ket{l} $ ($l=0,1,\ldots$) and the transition frequencies as $ \omega_{jk}\equiv \omega_{j}-\omega_{k}$, choosing the labeling of the states such that $\omega_{j}>\omega_{k}$, for $j>k$ [see Fig.~\ref{fig:5}(a)].
If the effective temperature of the mechanical reservoir is high enough to populate the state $\ket{5}  $, the system decays toward the ground state via two different one-photon  decay channels: $ \ket{5}\to \ket{2}\to \ket{0}  $ and $ \ket{3}\to \ket{2}\to \ket{0}  $. Since the  states $ \ket{5} $ and $ \ket{3} $ do not couple with the state $ \ket{4} $,  the other possible one-photon transition $  \ket{4}\to \ket{0}  $ can occur only by decays from  higher energy levels. 

We start considering the zero detuning case $ \Delta\equiv (\omega_{ c}-\omega_{ r})/\omega_{ m}=0 $, where $ \omega_{ r}\simeq1.017\, \omega_{ m} $ 
is the frequency corresponding to the minimum value of the splitting in Fig.~\ref{fig:2}(b). In this case the states $ \ket{3} $ and  $ \ket{5} $ are well approximated, respectively, by the superpositions $ \ket{\psi_{\pm}}=(\ket{0,2}\pm \ket{2,0_2})/\sqrt{2} $.

Figure~\ref{fig:5}(b) displays  the emission spectra for the system in the weak coupling regime, e.g., $ 2\Omega_{0,2}^{2,0}<\Gamma_{\rm tot} $. 
Due to the high value of $ \Gamma_{\rm tot} $, we observe a low-resolution emission spectrum  that only displays a wide band constituted  by a single peak at frequency $ \omega/\omega_m\simeq 0.98 $. On the contrary, when the system is in the strong coupling regime  ($ 2\Omega_{0,2}^{2,0}>\Gamma_{\rm tot} $), the spectrum becomes well-resolved. As shown in Fig.~\ref{fig:5}(b), for $ 2\Omega_{0,2}^{2,0}/\Gamma_{\rm tot}\simeq6.7 $  the cavity emission spectrum  displays two main peaks. Indeed, in the resonant case the accidentally quasi-degenerate transitions  $ \ket{5}\to \ket{2}  $ and $ \ket{2}\to \ket{0}  $  give rise to a single high-frequency peak at $ \omega\simeq\omega_m $, whereas the lower-frequency peak at $ \omega/\omega_m\simeq 0.98 $ corresponds to the transition $\ket{3}\to\ket{2} $.
It is important to notice that, in the presence of a $ T_{\kappa}=0 $  cavity reservoir, these peaks are observable only if the $ V_{\rm DCE}   $ term is included in the Hamiltonian. Indeed, without this term the states  $ \ket{2,0_2} $ and   $ \ket{0,2} $ are not coupled anymore and since the mechanical incoherent pumping only populates phononic states, the one-phonon decay peaks cannot be observed in the cavity emission spectra.  
We now turn  to the numerical analysis of the detuning effects on the cavity emission spectra. Figure~\ref{fig:6}(a) displays the emission  spectrum calculated for $ \Delta=0.028 $. As the transitions $ \ket{5}\to \ket{2}  $ and $ \ket{2}\to \ket{0}  $ are no more quasi-degenerate, the peaks at frequencies $ \omega_{52} $ and  $ \omega_{20}$ become well resolved, while the peak corresponding to $ \ket{3}\to \ket{2}  $ shifts towards a slightly lower frequency. As expected,  if we reduce the detuning and approach  the resonance point $ \Delta=0 $, the spectrum essentially  presents the same main features of Fig.~\ref{fig:5}(c). Specifically,  Fig.~\ref{fig:6}(b) shows that for $ \Delta=0.014 $ the two peaks at $ \omega_{52} $ and  $ \omega_{20}$ merge, and the emission spectrum presents only a main contibution at $ \omega/\omega_{ m}\simeq 1.015 $, while the transition frequency $ \omega_{32} $ does not change significantly.
Finally, in  Fig.~\ref{fig:6}(c) we study the emission spectrum in the presence of a negative detuning $ \Delta=-0.028 $. Also in this case, the spectrum displays three distinct peaks  placed at lower frequencies respect to Fig.~\ref{fig:6}(a). This shift arises from the energy-level crossing between the states $\ket{1} $ and $ \ket{2}$ shown in Fig.~\ref{fig:2}(a). Although the highest peak still corresponds to the one-photon decay toward the ground state, the emission spectrum is not symmetric respect to the one in Fig.~\ref{fig:6}(a). In particular, we observe that the intensity of the peak associated to the  transition $\ket{3}\to\ket{1} $ increases, whereas the $\ket{5}\to\ket{1} $ transition peak displays a much lower intensity. This effect can be explained considering that, differently from the positive-detuning cases studied above, for  $ \Delta<0 $ the state $ \ket{3}\simeq \ket{2,0_2} $ has more photonic character than $ \ket{5}\simeq \ket{0,2} $, which has more phononic character. Thus, while the photonic character of the polaron state $ \ket{3} $ leads to an enhancement of the  peak intensity at $ \omega_{31} $ in the cavity emission spectrum, on the other hand the phononic character of the state $ \ket{5} $ is responsible for the intensity decrease of the peak at $ \omega_{51} $. This study provides useful information on the emission process. Moreover, the presence of these features in the experimental spectra would represent a signature of the production of  DCE photons.

\section{Conclusions}
\label{sec:VI}
In conclusion, we have studied the dynamical Casimir effect in cavity optomechanics achieved only under incoherent mechanical excitation. We employed a fully quantum-mechanical description of both the cavity field and the oscillating mirror. The system dynamics is evaluated under incoherent pumping of the mechanical component, provided by a thermal-like excitation. Using a master equation approach in order to take into account losses, thermal effects and decoherence in the presence of a quasi-harmonic spectrum, we showed that a measurable flux of Casimir photons can be obtained also without a coherent pumping, suggesting another way for experimental observation of the DCE. The energy transfer of incoherent mechanical excitations to phonon pairs can be also observed in parametrically amplified optomechanical systems \cite{qin2019}. 
In Ref.~\cite{macri2018}, it has been shown that a vibrating mirror is affected by spontaneous emission in analogy with ordinary atoms. However, it decays emitting photon pairs. Here, we show that an incoherently excited vibrating mirror can emit light, in analogy to atomic fluorescence or electroluminescence in semiconductor devices. 

By applying the quantum regression theorem, we have calculated numerically the steady-state cavity emission spectra under incoherent mechanical excitation, for different detunings and loss rates. When the loss rates are lower than the effective coupling rate, the emission spectra allow to identify the different emission channels.

\acknowledgments{FN is supported in part by the: MURI Center for Dynamic Magneto-Optics via the Air Force Office of Scientific Research (AFOSR) (FA9550-14-1-0040), Army Research Office (ARO) (Grant No. W911NF-18-1-0358), Asian Office of Aerospace Research and Development (AOARD) (Grant No. FA2386-18-1-4045), Japan Science and Technology Agency (JST) (via the Q-LEAP program,the ImPACT program and CREST Grant No. JPMJCR1676), Japan Society for the Promotion of Science (JSPS) (JSPS-RFBR Grant No. 17-52-50023, and JSPS-FWO Grant No. VS.059.18N), RIKEN-AIST Challenge Research Fund, and the John Templeton Foundation. S.S. acknowledges the Army Research Office (ARO) (Grant No. W911NF1910065).}

\newpage
	\bibliography{DCE2}

\begin{thebibliography}{80}%
\makeatletter
\providecommand \@ifxundefined [1]{%
 \@ifx{#1\undefined}
}%
\providecommand \@ifnum [1]{%
 \ifnum #1\expandafter \@firstoftwo
 \else \expandafter \@secondoftwo
 \fi
}%
\providecommand \@ifx [1]{%
 \ifx #1\expandafter \@firstoftwo
 \else \expandafter \@secondoftwo
 \fi
}%
\providecommand \natexlab [1]{#1}%
\providecommand \enquote  [1]{``#1''}%
\providecommand \bibnamefont  [1]{#1}%
\providecommand \bibfnamefont [1]{#1}%
\providecommand \citenamefont [1]{#1}%
\providecommand \href@noop [0]{\@secondoftwo}%
\providecommand \href [0]{\begingroup \@sanitize@url \@href}%
\providecommand \@href[1]{\@@startlink{#1}\@@href}%
\providecommand \@@href[1]{\endgroup#1\@@endlink}%
\providecommand \@sanitize@url [0]{\catcode `\\12\catcode `\$12\catcode
  `\&12\catcode `\#12\catcode `\^12\catcode `\_12\catcode `\%12\relax}%
\providecommand \@@startlink[1]{}%
\providecommand \@@endlink[0]{}%
\providecommand \url  [0]{\begingroup\@sanitize@url \@url }%
\providecommand \@url [1]{\endgroup\@href {#1}{\urlprefix }}%
\providecommand \urlprefix  [0]{URL }%
\providecommand \Eprint [0]{\href }%
\providecommand \doibase [0]{http://dx.doi.org/}%
\providecommand \selectlanguage [0]{\@gobble}%
\providecommand \bibinfo  [0]{\@secondoftwo}%
\providecommand \bibfield  [0]{\@secondoftwo}%
\providecommand \translation [1]{[#1]}%
\providecommand \BibitemOpen [0]{}%
\providecommand \bibitemStop [0]{}%
\providecommand \bibitemNoStop [0]{.\EOS\space}%
\providecommand \EOS [0]{\spacefactor3000\relax}%
\providecommand \BibitemShut  [1]{\csname bibitem#1\endcsname}%
\let\auto@bib@innerbib\@empty
\bibitem [{\citenamefont {Schwinger}(1951)}]{Schwinger1951}%
  \BibitemOpen
  \bibfield  {author} {\bibinfo {author} {\bibfnamefont {J.}~\bibnamefont
  {Schwinger}},\ }\bibfield  {title} {\enquote {\bibinfo {title} {{On gauge
  invariance and vacuum polarization}},}\ }\href {\doibase
  10.1103/PhysRev.82.664} {\bibfield  {journal} {\bibinfo  {journal} {Am. J.
  Phys}\ }\textbf {\bibinfo {volume} {82}},\ \bibinfo {pages} {664--679}
  (\bibinfo {year} {1951})}\BibitemShut {NoStop}%
\bibitem [{\citenamefont {Moore}(1970)}]{Moore1970}%
  \BibitemOpen
  \bibfield  {author} {\bibinfo {author} {\bibfnamefont {G.~T.}\ \bibnamefont
  {Moore}},\ }\bibfield  {title} {\enquote {\bibinfo {title} {{Quantum theory
  of the electromagnetic field in a variable-length one-dimensional cavity}},}\
  }\href {\doibase 10.1063/1.1665432} {\bibfield  {journal} {\bibinfo
  {journal} {J. Math. Phys.}\ }\textbf {\bibinfo {volume} {11}},\ \bibinfo
  {pages} {2679} (\bibinfo {year} {1970})}\BibitemShut {NoStop}%
\bibitem [{\citenamefont {Fulling}\ and\ \citenamefont
  {Davies}(1976)}]{Fulling1976}%
  \BibitemOpen
  \bibfield  {author} {\bibinfo {author} {\bibfnamefont {S.~A.}\ \bibnamefont
  {Fulling}}\ and\ \bibinfo {author} {\bibfnamefont {P.~C.~W.}\ \bibnamefont
  {Davies}},\ }\bibfield  {title} {\enquote {\bibinfo {title} {{Radiation from
  a moving mirror in two dimensional space-time: conformal anomaly}},}\ }\href
  {https://royalsocietypublishing.org/doi/abs/10.1098/rspa.1976.0045}
  {\bibfield  {journal} {\bibinfo  {journal} {Proc. R. Soc. A.}\ }\textbf
  {\bibinfo {volume} {348}},\ \bibinfo {pages} {393--414} (\bibinfo {year}
  {1976})}\BibitemShut {NoStop}%
\bibitem [{\citenamefont {Yablonovitch}(1989)}]{Yablonovitch1989}%
  \BibitemOpen
  \bibfield  {author} {\bibinfo {author} {\bibfnamefont {E.}~\bibnamefont
  {Yablonovitch}},\ }\bibfield  {title} {\enquote {\bibinfo {title}
  {{Accelerating reference frame for electromagnetic waves in a rapidly growing
  plasma: {U}nruh-{D}avies-{F}ulling-{D}e{W}itt radiation and the nonadiabatic
  {C}asimir effect}},}\ }\href {\doibase 10.1103/PhysRevLett.62.1742}
  {\bibfield  {journal} {\bibinfo  {journal} {Phys. Rev. Lett.}\ }\textbf
  {\bibinfo {volume} {62}},\ \bibinfo {pages} {1742} (\bibinfo {year}
  {1989})}\BibitemShut {NoStop}%
\bibitem [{\citenamefont {Schwinger}(1993)}]{Schwinger1993}%
  \BibitemOpen
  \bibfield  {author} {\bibinfo {author} {\bibfnamefont {J.}~\bibnamefont
  {Schwinger}},\ }\bibfield  {title} {\enquote {\bibinfo {title} {{{C}asimir
  light: the source}},}\ }\href {\doibase 10.1073/pnas.90.6.2105} {\bibfield
  {journal} {\bibinfo  {journal} {PNAS}\ }\textbf {\bibinfo {volume} {90}},\
  \bibinfo {pages} {2105--2106} (\bibinfo {year} {1993})}\BibitemShut {NoStop}%
\bibitem [{\citenamefont {Scully}\ and\ \citenamefont
  {Zubairy}(1997)}]{Scully1997}%
  \BibitemOpen
  \bibfield  {author} {\bibinfo {author} {\bibfnamefont {M.~O.}\ \bibnamefont
  {Scully}}\ and\ \bibinfo {author} {\bibfnamefont {M.~S.}\ \bibnamefont
  {Zubairy}},\ }\href {\doibase doi:10.1017/CBO9780511813993} {\emph {\bibinfo
  {title} {{Quantum optics}}}}\ (\bibinfo  {publisher} {{Cambridge University
  Press}},\ \bibinfo {year} {1997})\BibitemShut {NoStop}%
\bibitem [{\citenamefont {Greiner}\ and\ \citenamefont
  {Schramm}(2008)}]{Greiner2008}%
  \BibitemOpen
  \bibfield  {author} {\bibinfo {author} {\bibfnamefont {W.}~\bibnamefont
  {Greiner}}\ and\ \bibinfo {author} {\bibfnamefont {S.}~\bibnamefont
  {Schramm}},\ }\bibfield  {title} {\enquote {\bibinfo {title} {{Resource
  letter QEDV-1: the QED vacuum}},}\ }\href {\doibase 10.1119/1.2820395}
  {\bibfield  {journal} {\bibinfo  {journal} {Am. J. Phys}\ }\textbf {\bibinfo
  {volume} {76}},\ \bibinfo {pages} {509--518} (\bibinfo {year}
  {2008})}\BibitemShut {NoStop}%
\bibitem [{\citenamefont {Dodonov}\ and\ \citenamefont
  {Klimov}(1992)}]{Dodonov1992}%
  \BibitemOpen
  \bibfield  {author} {\bibinfo {author} {\bibfnamefont {V.~V.}\ \bibnamefont
  {Dodonov}}\ and\ \bibinfo {author} {\bibfnamefont {A.~B.}\ \bibnamefont
  {Klimov}},\ }\bibfield  {title} {\enquote {\bibinfo {title} {{Long-time
  asymptotics of a quantized electromagnetic field in a resonator with
  oscillating boundary}},}\ }\href
  {https://www.sciencedirect.com/science/article/abs/pii/0375960192902125}
  {\bibfield  {journal} {\bibinfo  {journal} {Phys. Lett. A}\ }\textbf
  {\bibinfo {volume} {167}},\ \bibinfo {pages} {309--313} (\bibinfo {year}
  {1992})}\BibitemShut {NoStop}%
\bibitem [{\citenamefont {Dodonov}\ \emph {et~al.}(1993)\citenamefont
  {Dodonov}, \citenamefont {Klimov},\ and\ \citenamefont
  {Nikonov}}]{Dodonov1993}%
  \BibitemOpen
  \bibfield  {author} {\bibinfo {author} {\bibfnamefont {V.~V.}\ \bibnamefont
  {Dodonov}}, \bibinfo {author} {\bibfnamefont {A.~B.}\ \bibnamefont {Klimov}},
  \ and\ \bibinfo {author} {\bibfnamefont {D.~E.}\ \bibnamefont {Nikonov}},\
  }\bibfield  {title} {\enquote {\bibinfo {title} {{Quantum phenomena in
  nonstationary media}},}\ }\href
  {https://journals.aps.org/pra/abstract/10.1103/PhysRevA.47.4422} {\bibfield
  {journal} {\bibinfo  {journal} {Phys. Rev. A}\ }\textbf {\bibinfo {volume}
  {47}},\ \bibinfo {pages} {4422} (\bibinfo {year} {1993})}\BibitemShut
  {NoStop}%
\bibitem [{\citenamefont {Ji}\ \emph {et~al.}(1997)\citenamefont {Ji},
  \citenamefont {Jung}, \citenamefont {Park},\ and\ \citenamefont
  {Soh}}]{Ji1997}%
  \BibitemOpen
  \bibfield  {author} {\bibinfo {author} {\bibfnamefont {J.-Y.}\ \bibnamefont
  {Ji}}, \bibinfo {author} {\bibfnamefont {H.-H.}\ \bibnamefont {Jung}},
  \bibinfo {author} {\bibfnamefont {J.-W.}\ \bibnamefont {Park}}, \ and\
  \bibinfo {author} {\bibfnamefont {K.-S.}\ \bibnamefont {Soh}},\ }\bibfield
  {title} {\enquote {\bibinfo {title} {{Production of photons by the parametric
  resonance in the dynamical Casimir effect}},}\ }\href
  {https://journals.aps.org/pra/abstract/10.1103/PhysRevA.56.4440} {\bibfield
  {journal} {\bibinfo  {journal} {Phys. Rev. A}\ }\textbf {\bibinfo {volume}
  {56}},\ \bibinfo {pages} {4440} (\bibinfo {year} {1997})}\BibitemShut
  {NoStop}%
\bibitem [{\citenamefont {Mundarain}\ and\ \citenamefont {{Maia
  Neto}}(1998)}]{Mundarain1998}%
  \BibitemOpen
  \bibfield  {author} {\bibinfo {author} {\bibfnamefont {D.~F.}\ \bibnamefont
  {Mundarain}}\ and\ \bibinfo {author} {\bibfnamefont {P.~A.}\ \bibnamefont
  {{Maia Neto}}},\ }\bibfield  {title} {\enquote {\bibinfo {title} {{Quantum
  radiation in a plane cavity with moving mirrors}},}\ }\href {\doibase
  10.1103/PhysRevA.57.1379} {\bibfield  {journal} {\bibinfo  {journal} {Phys.
  Rev. A}\ }\textbf {\bibinfo {volume} {57}},\ \bibinfo {pages} {1379--1390}
  (\bibinfo {year} {1998})}\BibitemShut {NoStop}%
\bibitem [{\citenamefont {Dodonov}\ and\ \citenamefont
  {Mendon{\c{c}}a}(2014)}]{Dodonov2014}%
  \BibitemOpen
  \bibfield  {author} {\bibinfo {author} {\bibfnamefont {V.~V.}\ \bibnamefont
  {Dodonov}}\ and\ \bibinfo {author} {\bibfnamefont {J.~T.}\ \bibnamefont
  {Mendon{\c{c}}a}},\ }\bibfield  {title} {\enquote {\bibinfo {title}
  {{Dynamical Casimir effect in ultra-cold matter with a time-dependent
  effective charge}},}\ }\href
  {http://stacks.iop.org/1402-4896/2014/i=T160/a=014008} {\bibfield  {journal}
  {\bibinfo  {journal} {Phys. Scr.}\ }\textbf {\bibinfo {volume} {2014}},\
  \bibinfo {pages} {014008} (\bibinfo {year} {2014})}\BibitemShut {NoStop}%
\bibitem [{\citenamefont {Trautmann}\ and\ \citenamefont
  {Hauke}(2016)}]{Trautmann2016}%
  \BibitemOpen
  \bibfield  {author} {\bibinfo {author} {\bibfnamefont {N.}~\bibnamefont
  {Trautmann}}\ and\ \bibinfo {author} {\bibfnamefont {P.}~\bibnamefont
  {Hauke}},\ }\bibfield  {title} {\enquote {\bibinfo {title} {{Quantum
  simulation of the dynamical Casimir effect with trapped ions}},}\ }\href
  {http://stacks.iop.org/1367-2630/18/i=4/a=043029} {\bibfield  {journal}
  {\bibinfo  {journal} {New J. Phys.}\ }\textbf {\bibinfo {volume} {18}},\
  \bibinfo {pages} {043029} (\bibinfo {year} {2016})}\BibitemShut {NoStop}%
\bibitem [{\citenamefont {{Motazedifard}}\ \emph {et~al.}(2017)\citenamefont
  {{Motazedifard}}, \citenamefont {Naderi},\ and\ \citenamefont
  {Roknizadeh}}]{Motazedifard2017}%
  \BibitemOpen
  \bibfield  {author} {\bibinfo {author} {\bibfnamefont {A.}~\bibnamefont
  {{Motazedifard}}}, \bibinfo {author} {\bibfnamefont {M.~H.}\ \bibnamefont
  {Naderi}}, \ and\ \bibinfo {author} {\bibfnamefont {R.}~\bibnamefont
  {Roknizadeh}},\ }\bibfield  {title} {\enquote {\bibinfo {title} {{Dynamical
  Casimir effect of phonon excitation in the dispersive regime of cavity
  optomechanics}},}\ }\href {\doibase 10.1364/JOSAB.34.000642} {\bibfield
  {journal} {\bibinfo  {journal} {J. Opt. Soc. Am. B}\ }\textbf {\bibinfo
  {volume} {34}},\ \bibinfo {pages} {642--652} (\bibinfo {year}
  {2017})}\BibitemShut {NoStop}%
\bibitem [{\citenamefont {Carusotto}\ \emph {et~al.}(2010)\citenamefont
  {Carusotto}, \citenamefont {Balbinot}, \citenamefont {Fabbri},\ and\
  \citenamefont {Recati}}]{Carusotto2010}%
  \BibitemOpen
  \bibfield  {author} {\bibinfo {author} {\bibfnamefont {I.}~\bibnamefont
  {Carusotto}}, \bibinfo {author} {\bibfnamefont {R.}~\bibnamefont {Balbinot}},
  \bibinfo {author} {\bibfnamefont {A.}~\bibnamefont {Fabbri}}, \ and\ \bibinfo
  {author} {\bibfnamefont {A.}~\bibnamefont {Recati}},\ }\bibfield  {title}
  {\enquote {\bibinfo {title} {{Density correlations and analog dynamical
  Casimir emission of Bogoliubov phonons in modulated atomic Bose-Einstein
  condensates}},}\ }\href {\doibase 10.1140/epjd/e2009-00314-3} {\bibfield
  {journal} {\bibinfo  {journal} {Eur. Phys. J. D}\ }\textbf {\bibinfo {volume}
  {56}},\ \bibinfo {pages} {391--404} (\bibinfo {year} {2010})}\BibitemShut
  {NoStop}%
\bibitem [{\citenamefont {Jaskula}\ \emph {et~al.}(2012)\citenamefont
  {Jaskula}, \citenamefont {Partridge}, \citenamefont {Bonneau}, \citenamefont
  {Lopes}, \citenamefont {Ruaudel}, \citenamefont {Boiron},\ and\ \citenamefont
  {Westbrook}}]{Jaskula2012}%
  \BibitemOpen
  \bibfield  {author} {\bibinfo {author} {\bibfnamefont {J.-C.}\ \bibnamefont
  {Jaskula}}, \bibinfo {author} {\bibfnamefont {G.~B.}\ \bibnamefont
  {Partridge}}, \bibinfo {author} {\bibfnamefont {M.}~\bibnamefont {Bonneau}},
  \bibinfo {author} {\bibfnamefont {R.}~\bibnamefont {Lopes}}, \bibinfo
  {author} {\bibfnamefont {J.}~\bibnamefont {Ruaudel}}, \bibinfo {author}
  {\bibfnamefont {D.}~\bibnamefont {Boiron}}, \ and\ \bibinfo {author}
  {\bibfnamefont {C.~I.}\ \bibnamefont {Westbrook}},\ }\bibfield  {title}
  {\enquote {\bibinfo {title} {{Acoustic analog to the dynamical Casimir effect
  in a Bose-Einstein condensate}},}\ }\href {\doibase
  10.1103/PhysRevLett.109.220401} {\bibfield  {journal} {\bibinfo  {journal}
  {Phys. Rev. Lett.}\ }\textbf {\bibinfo {volume} {109}},\ \bibinfo {pages}
  {220401} (\bibinfo {year} {2012})}\BibitemShut {NoStop}%
\bibitem [{\citenamefont {{De Liberato}}\ \emph {et~al.}(2007)\citenamefont
  {{De Liberato}}, \citenamefont {Ciuti},\ and\ \citenamefont
  {Carusotto}}]{DeLiberato2007}%
  \BibitemOpen
  \bibfield  {author} {\bibinfo {author} {\bibfnamefont {S.}~\bibnamefont {{De
  Liberato}}}, \bibinfo {author} {\bibfnamefont {C.}~\bibnamefont {Ciuti}}, \
  and\ \bibinfo {author} {\bibfnamefont {I.}~\bibnamefont {Carusotto}},\
  }\bibfield  {title} {\enquote {\bibinfo {title} {{Quantum vacuum radiation
  spectra from a semiconductor microcavity with a time-modulated vacuum Rabi
  frequency}},}\ }\href {\doibase 10.1103/PhysRevLett.98.103602} {\bibfield
  {journal} {\bibinfo  {journal} {Phys. Rev. Lett.}\ }\textbf {\bibinfo
  {volume} {98}},\ \bibinfo {pages} {103602} (\bibinfo {year}
  {2007})}\BibitemShut {NoStop}%
\bibitem [{\citenamefont {Anappara}\ \emph {et~al.}(2009)\citenamefont
  {Anappara}, \citenamefont {De~Liberato}, \citenamefont {Tredicucci},
  \citenamefont {Ciuti}, \citenamefont {Biasiol}, \citenamefont {Sorba},\ and\
  \citenamefont {Beltram}}]{Anappara2009}%
  \BibitemOpen
  \bibfield  {author} {\bibinfo {author} {\bibfnamefont {A.~A.}\ \bibnamefont
  {Anappara}}, \bibinfo {author} {\bibfnamefont {S.}~\bibnamefont
  {De~Liberato}}, \bibinfo {author} {\bibfnamefont {A.}~\bibnamefont
  {Tredicucci}}, \bibinfo {author} {\bibfnamefont {C.}~\bibnamefont {Ciuti}},
  \bibinfo {author} {\bibfnamefont {G.}~\bibnamefont {Biasiol}}, \bibinfo
  {author} {\bibfnamefont {L.}~\bibnamefont {Sorba}}, \ and\ \bibinfo {author}
  {\bibfnamefont {F.}~\bibnamefont {Beltram}},\ }\bibfield  {title} {\enquote
  {\bibinfo {title} {{Signatures of the ultrastrong light-matter coupling
  regime}},}\ }\href {\doibase 10.1103/PhysRevB.79.201303} {\bibfield
  {journal} {\bibinfo  {journal} {Phys. Rev. B}\ }\textbf {\bibinfo {volume}
  {79}},\ \bibinfo {pages} {201303} (\bibinfo {year} {2009})}\BibitemShut
  {NoStop}%
\bibitem [{\citenamefont {De~Liberato}\ \emph {et~al.}(2009)\citenamefont
  {De~Liberato}, \citenamefont {Gerace}, \citenamefont {Carusotto},\ and\
  \citenamefont {Ciuti}}]{DeLiberato2009}%
  \BibitemOpen
  \bibfield  {author} {\bibinfo {author} {\bibfnamefont {S.}~\bibnamefont
  {De~Liberato}}, \bibinfo {author} {\bibfnamefont {D.}~\bibnamefont {Gerace}},
  \bibinfo {author} {\bibfnamefont {I.}~\bibnamefont {Carusotto}}, \ and\
  \bibinfo {author} {\bibfnamefont {C.}~\bibnamefont {Ciuti}},\ }\bibfield
  {title} {\enquote {\bibinfo {title} {{Extracavity quantum vacuum radiation
  from a single qubit}},}\ }\href {\doibase 10.1103/PhysRevA.80.053810}
  {\bibfield  {journal} {\bibinfo  {journal} {Phys. Rev. A}\ }\textbf {\bibinfo
  {volume} {80}},\ \bibinfo {pages} {053810} (\bibinfo {year}
  {2009})}\BibitemShut {NoStop}%
\bibitem [{\citenamefont {Garziano}\ \emph {et~al.}(2013)\citenamefont
  {Garziano}, \citenamefont {Ridolfo}, \citenamefont {Stassi}, \citenamefont
  {{Di Stefano}},\ and\ \citenamefont {Savasta}}]{Garziano2013}%
  \BibitemOpen
  \bibfield  {author} {\bibinfo {author} {\bibfnamefont {L.}~\bibnamefont
  {Garziano}}, \bibinfo {author} {\bibfnamefont {A.}~\bibnamefont {Ridolfo}},
  \bibinfo {author} {\bibfnamefont {R.}~\bibnamefont {Stassi}}, \bibinfo
  {author} {\bibfnamefont {O.}~\bibnamefont {{Di Stefano}}}, \ and\ \bibinfo
  {author} {\bibfnamefont {S.}~\bibnamefont {Savasta}},\ }\bibfield  {title}
  {\enquote {\bibinfo {title} {{Switching on and off of ultrastrong
  light-matter interaction: photon statistics of quantum vacuum radiation}},}\
  }\href {\doibase 10.1103/PhysRevA.88.063829} {\bibfield  {journal} {\bibinfo
  {journal} {Phys. Rev. A}\ }\textbf {\bibinfo {volume} {88}},\ \bibinfo
  {pages} {063829} (\bibinfo {year} {2013})}\BibitemShut {NoStop}%
\bibitem [{\citenamefont {Mu{\~n}oz}\ \emph {et~al.}(2018)\citenamefont
  {Mu{\~n}oz}, \citenamefont {Nori},\ and\ \citenamefont {{De
  Liberato}}}]{Munoz2018}%
  \BibitemOpen
  \bibfield  {author} {\bibinfo {author} {\bibfnamefont {C.~S.}\ \bibnamefont
  {Mu{\~n}oz}}, \bibinfo {author} {\bibfnamefont {F.}~\bibnamefont {Nori}}, \
  and\ \bibinfo {author} {\bibfnamefont {S.}~\bibnamefont {{De Liberato}}},\
  }\bibfield  {title} {\enquote {\bibinfo {title} {{Resolution of superluminal
  signalling in non-perturbative cavity quantum electrodynamics}},}\ }\href
  {https://www.nature.com/articles/s41467-018-04339-w} {\bibfield  {journal}
  {\bibinfo  {journal} {Nat. Commun.}\ }\textbf {\bibinfo {volume} {9}},\
  \bibinfo {pages} {1924} (\bibinfo {year} {2018})}\BibitemShut {NoStop}%
\bibitem [{\citenamefont {Lambrecht}\ \emph {et~al.}(1996)\citenamefont
  {Lambrecht}, \citenamefont {Jaekel},\ and\ \citenamefont
  {Reynaud}}]{Lambrecht1996}%
  \BibitemOpen
  \bibfield  {author} {\bibinfo {author} {\bibfnamefont {A.}~\bibnamefont
  {Lambrecht}}, \bibinfo {author} {\bibfnamefont {M.-T.}\ \bibnamefont
  {Jaekel}}, \ and\ \bibinfo {author} {\bibfnamefont {S.}~\bibnamefont
  {Reynaud}},\ }\bibfield  {title} {\enquote {\bibinfo {title} {{Motion induced
  radiation from a vibrating cavity}},}\ }\href {\doibase
  10.1103/PhysRevLett.77.615} {\bibfield  {journal} {\bibinfo  {journal} {Phys.
  Rev. Lett.}\ }\textbf {\bibinfo {volume} {77}},\ \bibinfo {pages} {615--618}
  (\bibinfo {year} {1996})}\BibitemShut {NoStop}%
\bibitem [{\citenamefont {Ford}\ and\ \citenamefont
  {Vilenkin}(1982)}]{Ford1982}%
  \BibitemOpen
  \bibfield  {author} {\bibinfo {author} {\bibfnamefont {L.~H.}\ \bibnamefont
  {Ford}}\ and\ \bibinfo {author} {\bibfnamefont {A.}~\bibnamefont
  {Vilenkin}},\ }\bibfield  {title} {\enquote {\bibinfo {title} {{Quantum
  radiation by moving mirrors}},}\ }\href {\doibase 10.1103/PhysRevD.25.2569}
  {\bibfield  {journal} {\bibinfo  {journal} {Phys. Rev. D}\ }\textbf {\bibinfo
  {volume} {25}},\ \bibinfo {pages} {2569--2575} (\bibinfo {year}
  {1982})}\BibitemShut {NoStop}%
\bibitem [{\citenamefont {Barton}\ and\ \citenamefont
  {Eberlein}(1993)}]{BARTON1993}%
  \BibitemOpen
  \bibfield  {author} {\bibinfo {author} {\bibfnamefont {G.}~\bibnamefont
  {Barton}}\ and\ \bibinfo {author} {\bibfnamefont {C.}~\bibnamefont
  {Eberlein}},\ }\bibfield  {title} {\enquote {\bibinfo {title} {{On quantum
  radiation from a moving body with finite refractive index}},}\ }\href
  {http://www.sciencedirect.com/science/article/pii/S000349168371081X}
  {\bibfield  {journal} {\bibinfo  {journal} {Ann. Phys.}\ }\textbf {\bibinfo
  {volume} {227}},\ \bibinfo {pages} {222 -- 274} (\bibinfo {year}
  {1993})}\BibitemShut {NoStop}%
\bibitem [{\citenamefont {Sassaroli}\ \emph {et~al.}(1994)\citenamefont
  {Sassaroli}, \citenamefont {Srivastava},\ and\ \citenamefont
  {Widom}}]{Sassaroli1994}%
  \BibitemOpen
  \bibfield  {author} {\bibinfo {author} {\bibfnamefont {E.}~\bibnamefont
  {Sassaroli}}, \bibinfo {author} {\bibfnamefont {Y.~N.}\ \bibnamefont
  {Srivastava}}, \ and\ \bibinfo {author} {\bibfnamefont {A.}~\bibnamefont
  {Widom}},\ }\bibfield  {title} {\enquote {\bibinfo {title} {{Photon
  production by the dynamical {C}asimir effect}},}\ }\href {\doibase
  10.1103/PhysRevA.50.1027} {\bibfield  {journal} {\bibinfo  {journal} {Phys.
  Rev. A}\ }\textbf {\bibinfo {volume} {50}},\ \bibinfo {pages} {1027--1034}
  (\bibinfo {year} {1994})}\BibitemShut {NoStop}%
\bibitem [{\citenamefont {Dodonov}\ and\ \citenamefont
  {Klimov}(1996)}]{Dodonov1996}%
  \BibitemOpen
  \bibfield  {author} {\bibinfo {author} {\bibfnamefont {V.~V.}\ \bibnamefont
  {Dodonov}}\ and\ \bibinfo {author} {\bibfnamefont {A.~B.}\ \bibnamefont
  {Klimov}},\ }\bibfield  {title} {\enquote {\bibinfo {title} {{Generation and
  detection of photons in a cavity with a resonantly oscillating boundary}},}\
  }\href {\doibase 10.1103/PhysRevA.53.2664} {\bibfield  {journal} {\bibinfo
  {journal} {Phys. Rev. A}\ }\textbf {\bibinfo {volume} {53}},\ \bibinfo
  {pages} {2664--2682} (\bibinfo {year} {1996})}\BibitemShut {NoStop}%
\bibitem [{\citenamefont {Schaller}\ \emph {et~al.}(2002)\citenamefont
  {Schaller}, \citenamefont {Sch\"utzhold}, \citenamefont {Plunien},\ and\
  \citenamefont {Soff}}]{Schaller2002}%
  \BibitemOpen
  \bibfield  {author} {\bibinfo {author} {\bibfnamefont {G.}~\bibnamefont
  {Schaller}}, \bibinfo {author} {\bibfnamefont {R.}~\bibnamefont
  {Sch\"utzhold}}, \bibinfo {author} {\bibfnamefont {G.}~\bibnamefont
  {Plunien}}, \ and\ \bibinfo {author} {\bibfnamefont {G.}~\bibnamefont
  {Soff}},\ }\bibfield  {title} {\enquote {\bibinfo {title} {{Dynamical
  {C}asimir effect in a leaky cavity at finite temperature}},}\ }\href
  {\doibase 10.1103/PhysRevA.66.023812} {\bibfield  {journal} {\bibinfo
  {journal} {Phys. Rev. A}\ }\textbf {\bibinfo {volume} {66}},\ \bibinfo
  {pages} {023812} (\bibinfo {year} {2002})}\BibitemShut {NoStop}%
\bibitem [{\citenamefont {Kim}\ \emph {et~al.}(2006)\citenamefont {Kim},
  \citenamefont {Brownell},\ and\ \citenamefont {Onofrio}}]{Kim2006}%
  \BibitemOpen
  \bibfield  {author} {\bibinfo {author} {\bibfnamefont {W.-J.}\ \bibnamefont
  {Kim}}, \bibinfo {author} {\bibfnamefont {J.~H.}\ \bibnamefont {Brownell}}, \
  and\ \bibinfo {author} {\bibfnamefont {R.}~\bibnamefont {Onofrio}},\
  }\bibfield  {title} {\enquote {\bibinfo {title} {{Detectability of
  dissipative motion in quantum vacuum via superradiance}},}\ }\href {\doibase
  10.1103/PhysRevLett.96.200402} {\bibfield  {journal} {\bibinfo  {journal}
  {Phys. Rev. Lett.}\ }\textbf {\bibinfo {volume} {96}},\ \bibinfo {pages}
  {200402} (\bibinfo {year} {2006})}\BibitemShut {NoStop}%
\bibitem [{\citenamefont {Dodonov}(2010)}]{Dodonov2010}%
  \BibitemOpen
  \bibfield  {author} {\bibinfo {author} {\bibfnamefont {V.~V.}\ \bibnamefont
  {Dodonov}},\ }\bibfield  {title} {\enquote {\bibinfo {title} {{Current status
  of the dynamical {C}asimir effect}},}\ }\href {\doibase
  10.1088/0031-8949/82/03/038105} {\bibfield  {journal} {\bibinfo  {journal}
  {Phys. Scr.}\ }\textbf {\bibinfo {volume} {82}},\ \bibinfo {pages} {038105}
  (\bibinfo {year} {2010})}\BibitemShut {NoStop}%
\bibitem [{\citenamefont {Lozovik}\ \emph {et~al.}(1995)\citenamefont
  {Lozovik}, \citenamefont {Tsvetus},\ and\ \citenamefont
  {Vinogradov}}]{Lozovik1995}%
  \BibitemOpen
  \bibfield  {author} {\bibinfo {author} {\bibfnamefont {Y.~E.}\ \bibnamefont
  {Lozovik}}, \bibinfo {author} {\bibfnamefont {V.~G.}\ \bibnamefont
  {Tsvetus}}, \ and\ \bibinfo {author} {\bibfnamefont {E.~A.}\ \bibnamefont
  {Vinogradov}},\ }\bibfield  {title} {\enquote {\bibinfo {title} {{Femtosecond
  parametric excitation of electromagnetic field in a cavity}},}\ }\href
  {http://jetpletters.ac.ru/ps/1208/article_18257.pdf} {\bibfield  {journal}
  {\bibinfo  {journal} {JETP Lett.}\ }\textbf {\bibinfo {volume} {61}},\
  \bibinfo {pages} {723} (\bibinfo {year} {1995})}\BibitemShut {NoStop}%
\bibitem [{\citenamefont {Uhlmann}\ \emph {et~al.}(2004)\citenamefont
  {Uhlmann}, \citenamefont {Plunien}, \citenamefont {Sch\"utzhold},\ and\
  \citenamefont {Soff}}]{Uhlmann2004}%
  \BibitemOpen
  \bibfield  {author} {\bibinfo {author} {\bibfnamefont {M.}~\bibnamefont
  {Uhlmann}}, \bibinfo {author} {\bibfnamefont {G.}~\bibnamefont {Plunien}},
  \bibinfo {author} {\bibfnamefont {R.}~\bibnamefont {Sch\"utzhold}}, \ and\
  \bibinfo {author} {\bibfnamefont {G.}~\bibnamefont {Soff}},\ }\bibfield
  {title} {\enquote {\bibinfo {title} {{Resonant cavity photon creation via the
  dynamical {C}asimir effect}},}\ }\href {\doibase
  10.1103/PhysRevLett.93.193601} {\bibfield  {journal} {\bibinfo  {journal}
  {Phys. Rev. Lett.}\ }\textbf {\bibinfo {volume} {93}},\ \bibinfo {pages}
  {193601} (\bibinfo {year} {2004})}\BibitemShut {NoStop}%
\bibitem [{\citenamefont {Crocce}\ \emph {et~al.}(2004)\citenamefont {Crocce},
  \citenamefont {Dalvit}, \citenamefont {Lombardo},\ and\ \citenamefont
  {Mazzitelli}}]{Crocce2004}%
  \BibitemOpen
  \bibfield  {author} {\bibinfo {author} {\bibfnamefont {M.}~\bibnamefont
  {Crocce}}, \bibinfo {author} {\bibfnamefont {D.~A.~R.}\ \bibnamefont
  {Dalvit}}, \bibinfo {author} {\bibfnamefont {F.~C.}\ \bibnamefont
  {Lombardo}}, \ and\ \bibinfo {author} {\bibfnamefont {F.~D.}\ \bibnamefont
  {Mazzitelli}},\ }\bibfield  {title} {\enquote {\bibinfo {title} {{Model for
  resonant photon creation in a cavity with time-dependent conductivity}},}\
  }\href {\doibase 10.1103/PhysRevA.70.033811} {\bibfield  {journal} {\bibinfo
  {journal} {Phys. Rev. A}\ }\textbf {\bibinfo {volume} {70}},\ \bibinfo
  {pages} {033811} (\bibinfo {year} {2004})}\BibitemShut {NoStop}%
\bibitem [{\citenamefont {Braggio}\ \emph {et~al.}(2005)\citenamefont
  {Braggio}, \citenamefont {Bressi}, \citenamefont {Carugno}, \citenamefont
  {Noce}, \citenamefont {Galeazzi}, \citenamefont {Lombardi}, \citenamefont
  {Palmieri}, \citenamefont {Ruoso},\ and\ \citenamefont
  {Zanello}}]{Braggio2005}%
  \BibitemOpen
  \bibfield  {author} {\bibinfo {author} {\bibfnamefont {C.}~\bibnamefont
  {Braggio}}, \bibinfo {author} {\bibfnamefont {G.}~\bibnamefont {Bressi}},
  \bibinfo {author} {\bibfnamefont {G.}~\bibnamefont {Carugno}}, \bibinfo
  {author} {\bibfnamefont {C.~Del}\ \bibnamefont {Noce}}, \bibinfo {author}
  {\bibfnamefont {G.}~\bibnamefont {Galeazzi}}, \bibinfo {author}
  {\bibfnamefont {A.}~\bibnamefont {Lombardi}}, \bibinfo {author}
  {\bibfnamefont {A.}~\bibnamefont {Palmieri}}, \bibinfo {author}
  {\bibfnamefont {G.}~\bibnamefont {Ruoso}}, \ and\ \bibinfo {author}
  {\bibfnamefont {D.}~\bibnamefont {Zanello}},\ }\bibfield  {title} {\enquote
  {\bibinfo {title} {{A novel experimental approach for the detection of the
  dynamical {C}asimir effect}},}\ }\href {\doibase 10.1209/epl/i2005-10048-8}
  {\bibfield  {journal} {\bibinfo  {journal} {EPL}\ }\textbf {\bibinfo {volume}
  {70}},\ \bibinfo {pages} {754} (\bibinfo {year} {2005})}\BibitemShut
  {NoStop}%
\bibitem [{\citenamefont {Segev}\ \emph {et~al.}(2007)\citenamefont {Segev},
  \citenamefont {Abdo}, \citenamefont {Shtempluck}, \citenamefont {Buks},\ and\
  \citenamefont {Yurke}}]{Segev2007}%
  \BibitemOpen
  \bibfield  {author} {\bibinfo {author} {\bibfnamefont {E.}~\bibnamefont
  {Segev}}, \bibinfo {author} {\bibfnamefont {B.}~\bibnamefont {Abdo}},
  \bibinfo {author} {\bibfnamefont {O.}~\bibnamefont {Shtempluck}}, \bibinfo
  {author} {\bibfnamefont {E.}~\bibnamefont {Buks}}, \ and\ \bibinfo {author}
  {\bibfnamefont {B.}~\bibnamefont {Yurke}},\ }\bibfield  {title} {\enquote
  {\bibinfo {title} {{Prospects of employing superconducting stripline
  resonators for studying the dynamical {C}asimir effect experimentally}},}\
  }\href {http://www.sciencedirect.com/science/article/pii/S0375960107008067}
  {\bibfield  {journal} {\bibinfo  {journal} {Phys. Lett. A}\ }\textbf
  {\bibinfo {volume} {370}},\ \bibinfo {pages} {202--206} (\bibinfo {year}
  {2007})}\BibitemShut {NoStop}%
\bibitem [{\citenamefont {{De Melo e Souza}}\ \emph {et~al.}(2018)\citenamefont
  {{De Melo e Souza}}, \citenamefont {Impens},\ and\ \citenamefont {{Maia
  Neto}}}]{Souza2018}%
  \BibitemOpen
  \bibfield  {author} {\bibinfo {author} {\bibfnamefont {R.}~\bibnamefont {{De
  Melo e Souza}}}, \bibinfo {author} {\bibfnamefont {F.}~\bibnamefont
  {Impens}}, \ and\ \bibinfo {author} {\bibfnamefont {P.~A.}\ \bibnamefont
  {{Maia Neto}}},\ }\bibfield  {title} {\enquote {\bibinfo {title}
  {{Microscopic dynamical Casimir effect}},}\ }\href
  {https://journals.aps.org/pra/abstract/10.1103/PhysRevA.97.032514} {\bibfield
   {journal} {\bibinfo  {journal} {Phys. Rev. A}\ }\textbf {\bibinfo {volume}
  {97}},\ \bibinfo {pages} {032514} (\bibinfo {year} {2018})}\BibitemShut
  {NoStop}%
\bibitem [{\citenamefont {Johansson}\ \emph {et~al.}(2009)\citenamefont
  {Johansson}, \citenamefont {Johansson}, \citenamefont {Wilson},\ and\
  \citenamefont {Nori}}]{Johansson2009}%
  \BibitemOpen
  \bibfield  {author} {\bibinfo {author} {\bibfnamefont {J.~R.}\ \bibnamefont
  {Johansson}}, \bibinfo {author} {\bibfnamefont {G.}~\bibnamefont
  {Johansson}}, \bibinfo {author} {\bibfnamefont {C.~M.}\ \bibnamefont
  {Wilson}}, \ and\ \bibinfo {author} {\bibfnamefont {F.}~\bibnamefont
  {Nori}},\ }\bibfield  {title} {\enquote {\bibinfo {title} {{Dynamical
  {C}asimir effect in a superconducting coplanar waveguide}},}\ }\href
  {\doibase 10.1103/PhysRevLett.103.147003} {\bibfield  {journal} {\bibinfo
  {journal} {Phys. Rev. Lett.}\ }\textbf {\bibinfo {volume} {103}},\ \bibinfo
  {pages} {147003} (\bibinfo {year} {2009})}\BibitemShut {NoStop}%
\bibitem [{\citenamefont {Johansson}\ \emph {et~al.}(2010)\citenamefont
  {Johansson}, \citenamefont {Johansson}, \citenamefont {Wilson},\ and\
  \citenamefont {Nori}}]{Johansson2010}%
  \BibitemOpen
  \bibfield  {author} {\bibinfo {author} {\bibfnamefont {J.~R.}\ \bibnamefont
  {Johansson}}, \bibinfo {author} {\bibfnamefont {G.}~\bibnamefont
  {Johansson}}, \bibinfo {author} {\bibfnamefont {C.~M.}\ \bibnamefont
  {Wilson}}, \ and\ \bibinfo {author} {\bibfnamefont {F.}~\bibnamefont
  {Nori}},\ }\bibfield  {title} {\enquote {\bibinfo {title} {{Dynamical
  {C}asimir effect in superconducting microwave circuits}},}\ }\href {\doibase
  10.1103/PhysRevA.82.052509} {\bibfield  {journal} {\bibinfo  {journal} {Phys.
  Rev. A}\ }\textbf {\bibinfo {volume} {82}},\ \bibinfo {pages} {52509}
  (\bibinfo {year} {2010})}\BibitemShut {NoStop}%
\bibitem [{\citenamefont {Wilson}\ \emph {et~al.}(2011)\citenamefont {Wilson},
  \citenamefont {Johansson}, \citenamefont {Pourkabirian}, \citenamefont
  {Simoen}, \citenamefont {Johansson}, \citenamefont {Duty}, \citenamefont
  {Nori},\ and\ \citenamefont {Delsing}}]{Wilson2011}%
  \BibitemOpen
  \bibfield  {author} {\bibinfo {author} {\bibfnamefont {C.~M.}\ \bibnamefont
  {Wilson}}, \bibinfo {author} {\bibfnamefont {G.}~\bibnamefont {Johansson}},
  \bibinfo {author} {\bibfnamefont {A.}~\bibnamefont {Pourkabirian}}, \bibinfo
  {author} {\bibfnamefont {M.}~\bibnamefont {Simoen}}, \bibinfo {author}
  {\bibfnamefont {J.~R.}\ \bibnamefont {Johansson}}, \bibinfo {author}
  {\bibfnamefont {T.}~\bibnamefont {Duty}}, \bibinfo {author} {\bibfnamefont
  {F.}~\bibnamefont {Nori}}, \ and\ \bibinfo {author} {\bibfnamefont
  {P.}~\bibnamefont {Delsing}},\ }\bibfield  {title} {\enquote {\bibinfo
  {title} {{Observation of the dynamical {C}asimir effect in a superconducting
  circuit}},}\ }\href {\doibase 10.1038/nature10561} {\bibfield  {journal}
  {\bibinfo  {journal} {Nature}\ }\textbf {\bibinfo {volume} {479}},\ \bibinfo
  {pages} {376} (\bibinfo {year} {2011})}\BibitemShut {NoStop}%
\bibitem [{\citenamefont {Johansson}\ \emph {et~al.}(2013)\citenamefont
  {Johansson}, \citenamefont {Johansson}, \citenamefont {Wilson}, \citenamefont
  {Delsing},\ and\ \citenamefont {Nori}}]{Johansson2013}%
  \BibitemOpen
  \bibfield  {author} {\bibinfo {author} {\bibfnamefont {J.~R.}\ \bibnamefont
  {Johansson}}, \bibinfo {author} {\bibfnamefont {G.}~\bibnamefont
  {Johansson}}, \bibinfo {author} {\bibfnamefont {C.~M.}\ \bibnamefont
  {Wilson}}, \bibinfo {author} {\bibfnamefont {P.}~\bibnamefont {Delsing}}, \
  and\ \bibinfo {author} {\bibfnamefont {F.}~\bibnamefont {Nori}},\ }\bibfield
  {title} {\enquote {\bibinfo {title} {{Nonclassical microwave radiation from
  the dynamical {C}asimir effect}},}\ }\href {\doibase
  10.1103/PhysRevA.87.043804} {\bibfield  {journal} {\bibinfo  {journal} {Phys.
  Rev. A}\ }\textbf {\bibinfo {volume} {87}},\ \bibinfo {pages} {043804}
  (\bibinfo {year} {2013})}\BibitemShut {NoStop}%
\bibitem [{\citenamefont {Nation}\ \emph {et~al.}(2012)\citenamefont {Nation},
  \citenamefont {Johansson}, \citenamefont {Blencowe},\ and\ \citenamefont
  {Nori}}]{Nation2012}%
  \BibitemOpen
  \bibfield  {author} {\bibinfo {author} {\bibfnamefont {P.~D.}\ \bibnamefont
  {Nation}}, \bibinfo {author} {\bibfnamefont {J.~R.}\ \bibnamefont
  {Johansson}}, \bibinfo {author} {\bibfnamefont {M.~P.}\ \bibnamefont
  {Blencowe}}, \ and\ \bibinfo {author} {\bibfnamefont {F.}~\bibnamefont
  {Nori}},\ }\bibfield  {title} {\enquote {\bibinfo {title} {{Colloquium:
  stimulating uncertainty: amplifying the quantum vacuum with superconducting
  circuits}},}\ }\href {\doibase 10.1103/RevModPhys.84.1} {\bibfield  {journal}
  {\bibinfo  {journal} {Rev. Mod. Phys.}\ }\textbf {\bibinfo {volume} {84}},\
  \bibinfo {pages} {1--24} (\bibinfo {year} {2012})}\BibitemShut {NoStop}%
\bibitem [{\citenamefont {L{\"a}hteenm{\"a}ki}\ \emph
  {et~al.}(2013)\citenamefont {L{\"a}hteenm{\"a}ki}, \citenamefont {Paraoanu},
  \citenamefont {Hassel},\ and\ \citenamefont {Hakonen}}]{Laehteenmaeki2013}%
  \BibitemOpen
  \bibfield  {author} {\bibinfo {author} {\bibfnamefont {P.}~\bibnamefont
  {L{\"a}hteenm{\"a}ki}}, \bibinfo {author} {\bibfnamefont {G.~S.}\
  \bibnamefont {Paraoanu}}, \bibinfo {author} {\bibfnamefont {J.}~\bibnamefont
  {Hassel}}, \ and\ \bibinfo {author} {\bibfnamefont {P.~J.}\ \bibnamefont
  {Hakonen}},\ }\bibfield  {title} {\enquote {\bibinfo {title} {{Dynamical
  {C}asimir effect in a {J}osephson metamaterial}},}\ }\href {\doibase
  10.1073/pnas.1212705110} {\bibfield  {journal} {\bibinfo  {journal} {Proc.
  Natl. Acad. Sci. USA}\ }\textbf {\bibinfo {volume} {110}},\ \bibinfo {pages}
  {4234--4238} (\bibinfo {year} {2013})}\BibitemShut {NoStop}%
\bibitem [{\citenamefont {Law}(1995)}]{Law1995}%
  \BibitemOpen
  \bibfield  {author} {\bibinfo {author} {\bibfnamefont {C.~K.}\ \bibnamefont
  {Law}},\ }\bibfield  {title} {\enquote {\bibinfo {title} {{Interaction
  between a moving mirror and radiation pressure: a hamiltonian
  formulation}},}\ }\href {\doibase 10.1103/PhysRevA.51.2537} {\bibfield
  {journal} {\bibinfo  {journal} {Phys. Rev. A}\ }\textbf {\bibinfo {volume}
  {51}},\ \bibinfo {pages} {2537--2541} (\bibinfo {year} {1995})}\BibitemShut
  {NoStop}%
\bibitem [{\citenamefont {Sala}\ and\ \citenamefont
  {Tufarelli}(2018)}]{Sala2018}%
  \BibitemOpen
  \bibfield  {author} {\bibinfo {author} {\bibfnamefont {K.}~\bibnamefont
  {Sala}}\ and\ \bibinfo {author} {\bibfnamefont {T.}~\bibnamefont
  {Tufarelli}},\ }\bibfield  {title} {\enquote {\bibinfo {title} {{Exploring
  corrections to the Optomechanical Hamiltonian}},}\ }\href
  {https://www.nature.com/articles/s41598-018-26739-0} {\bibfield  {journal}
  {\bibinfo  {journal} {Sci. Rep.}\ }\textbf {\bibinfo {volume} {8}},\ \bibinfo
  {pages} {9157} (\bibinfo {year} {2018})}\BibitemShut {NoStop}%
\bibitem [{\citenamefont {Macr\`{\i}}\ \emph {et~al.}(2018)\citenamefont
  {Macr\`{\i}}, \citenamefont {Ridolfo}, \citenamefont {Di~Stefano},
  \citenamefont {Kockum}, \citenamefont {Nori},\ and\ \citenamefont
  {Savasta}}]{macri2018}%
  \BibitemOpen
  \bibfield  {author} {\bibinfo {author} {\bibfnamefont {V.}~\bibnamefont
  {Macr\`{\i}}}, \bibinfo {author} {\bibfnamefont {A.}~\bibnamefont {Ridolfo}},
  \bibinfo {author} {\bibfnamefont {O.}~\bibnamefont {Di~Stefano}}, \bibinfo
  {author} {\bibfnamefont {A.~F.}\ \bibnamefont {Kockum}}, \bibinfo {author}
  {\bibfnamefont {F.}~\bibnamefont {Nori}}, \ and\ \bibinfo {author}
  {\bibfnamefont {S.}~\bibnamefont {Savasta}},\ }\bibfield  {title} {\enquote
  {\bibinfo {title} {{Nonperturbative dynamical {C}asimir effect in
  optomechanical systems: vacuum {C}asimir-{R}abi splittings}},}\ }\href
  {\doibase 10.1103/PhysRevX.8.011031} {\bibfield  {journal} {\bibinfo
  {journal} {Phys. Rev. X}\ }\textbf {\bibinfo {volume} {8}},\ \bibinfo {pages}
  {011031} (\bibinfo {year} {2018})}\BibitemShut {NoStop}%
\bibitem [{\citenamefont {{Di Stefano}}\ \emph {et~al.}(2019)\citenamefont {{Di
  Stefano}}, \citenamefont {Settineri}, \citenamefont {Macr{\`{i}}},
  \citenamefont {Ridolfo}, \citenamefont {Stassi}, \citenamefont {Kockum},
  \citenamefont {Savasta},\ and\ \citenamefont {Nori}}]{DiStefano2019a}%
  \BibitemOpen
  \bibfield  {author} {\bibinfo {author} {\bibfnamefont {O.}~\bibnamefont {{Di
  Stefano}}}, \bibinfo {author} {\bibfnamefont {A.}~\bibnamefont {Settineri}},
  \bibinfo {author} {\bibfnamefont {V.}~\bibnamefont {Macr{\`{i}}}}, \bibinfo
  {author} {\bibfnamefont {A.}~\bibnamefont {Ridolfo}}, \bibinfo {author}
  {\bibfnamefont {R.}~\bibnamefont {Stassi}}, \bibinfo {author} {\bibfnamefont
  {A.~F.}\ \bibnamefont {Kockum}}, \bibinfo {author} {\bibfnamefont
  {S.}~\bibnamefont {Savasta}}, \ and\ \bibinfo {author} {\bibfnamefont
  {F.}~\bibnamefont {Nori}},\ }\bibfield  {title} {\enquote {\bibinfo {title}
  {{Interaction of mechanical oscillators mediated by the exchange of virtual
  photon pairs}},}\ }\href
  {https://journals.aps.org/prl/abstract/10.1103/PhysRevLett.122.030402}
  {\bibfield  {journal} {\bibinfo  {journal} {Phys. Rev. Lett.,}\ }\textbf
  {\bibinfo {volume} {122}},\ \bibinfo {pages} {030402} (\bibinfo {year}
  {2019})}\BibitemShut {NoStop}%
\bibitem [{\citenamefont {Wang}\ \emph {et~al.}(2018)\citenamefont {Wang},
  \citenamefont {Blencowe}, \citenamefont {Wilson},\ and\ \citenamefont
  {Rimberg}}]{Wang2018}%
  \BibitemOpen
  \bibfield  {author} {\bibinfo {author} {\bibfnamefont {H.~M.}\ \bibnamefont
  {Wang}}, \bibinfo {author} {\bibfnamefont {M.~P.}\ \bibnamefont {Blencowe}},
  \bibinfo {author} {\bibfnamefont {C.~M.}\ \bibnamefont {Wilson}}, \ and\
  \bibinfo {author} {\bibfnamefont {A.~J.}\ \bibnamefont {Rimberg}},\
  }\bibfield  {title} {\enquote {\bibinfo {title} {{Mechanically generating
  entangled photons from the vacuum: a microwave circuit-acoustic resonator
  analogue of the Unruh effect}},}\ }\href {https://arxiv.org/abs/1811.10065}
  {\  (\bibinfo {year} {2018})},\ \Eprint {http://arxiv.org/abs/1811.10065}
  {arXiv:1811.10065} \BibitemShut {NoStop}%
\bibitem [{\citenamefont {Kippenberg}\ and\ \citenamefont
  {Vahala}(2008)}]{Kippenberg1172}%
  \BibitemOpen
  \bibfield  {author} {\bibinfo {author} {\bibfnamefont {T.~J.}\ \bibnamefont
  {Kippenberg}}\ and\ \bibinfo {author} {\bibfnamefont {K.~J.}\ \bibnamefont
  {Vahala}},\ }\bibfield  {title} {\enquote {\bibinfo {title} {{Cavity
  optomechanics: back-action at the mesoscale}},}\ }\href {\doibase
  10.1126/science.1156032} {\bibfield  {journal} {\bibinfo  {journal}
  {Science}\ }\textbf {\bibinfo {volume} {321}},\ \bibinfo {pages} {1172--1176}
  (\bibinfo {year} {2008})}\BibitemShut {NoStop}%
\bibitem [{\citenamefont {Nunnenkamp}\ \emph {et~al.}(2011)\citenamefont
  {Nunnenkamp}, \citenamefont {B\o{}rkje},\ and\ \citenamefont
  {Girvin}}]{Nunnenkamp2011}%
  \BibitemOpen
  \bibfield  {author} {\bibinfo {author} {\bibfnamefont {A.}~\bibnamefont
  {Nunnenkamp}}, \bibinfo {author} {\bibfnamefont {K.}~\bibnamefont
  {B\o{}rkje}}, \ and\ \bibinfo {author} {\bibfnamefont {S.~M.}\ \bibnamefont
  {Girvin}},\ }\bibfield  {title} {\enquote {\bibinfo {title} {{Single-photon
  optomechanics}},}\ }\href {\doibase 10.1103/PhysRevLett.107.063602}
  {\bibfield  {journal} {\bibinfo  {journal} {Phys. Rev. Lett.}\ }\textbf
  {\bibinfo {volume} {107}},\ \bibinfo {pages} {063602} (\bibinfo {year}
  {2011})}\BibitemShut {NoStop}%
\bibitem [{\citenamefont {Marquardt}\ and\ \citenamefont
  {Girvin}(2009)}]{Marquardt2009}%
  \BibitemOpen
  \bibfield  {author} {\bibinfo {author} {\bibfnamefont {F.}~\bibnamefont
  {Marquardt}}\ and\ \bibinfo {author} {\bibfnamefont {S.~M.}\ \bibnamefont
  {Girvin}},\ }\bibfield  {title} {\enquote {\bibinfo {title}
  {{Optomechanics}},}\ }\href
  {https://physics.aps.org/articles/pdf/10.1103/Physics.2.40} {\bibfield
  {journal} {\bibinfo  {journal} {Physics}\ }\textbf {\bibinfo {volume} {2}},\
  \bibinfo {pages} {40} (\bibinfo {year} {2009})}\BibitemShut {NoStop}%
\bibitem [{\citenamefont {Teufel}\ \emph {et~al.}(2011)\citenamefont {Teufel},
  \citenamefont {Li}, \citenamefont {Allman}, \citenamefont {Cicak},
  \citenamefont {Sirois}, \citenamefont {Whittaker},\ and\ \citenamefont
  {Simmonds}}]{Teufel2011}%
  \BibitemOpen
  \bibfield  {author} {\bibinfo {author} {\bibfnamefont {J.~D.}\ \bibnamefont
  {Teufel}}, \bibinfo {author} {\bibfnamefont {D.}~\bibnamefont {Li}}, \bibinfo
  {author} {\bibfnamefont {M.~S.}\ \bibnamefont {Allman}}, \bibinfo {author}
  {\bibfnamefont {K.}~\bibnamefont {Cicak}}, \bibinfo {author} {\bibfnamefont
  {A.~J.}\ \bibnamefont {Sirois}}, \bibinfo {author} {\bibfnamefont {J.~D.}\
  \bibnamefont {Whittaker}}, \ and\ \bibinfo {author} {\bibfnamefont {R.~W.}\
  \bibnamefont {Simmonds}},\ }\bibfield  {title} {\enquote {\bibinfo {title}
  {{Circuit cavity electromechanics in the strong-coupling regime}},}\ }\href
  {\doibase 10.1038/nature09898} {\bibfield  {journal} {\bibinfo  {journal}
  {Nature}\ }\textbf {\bibinfo {volume} {471}},\ \bibinfo {pages} {204--208}
  (\bibinfo {year} {2011})}\BibitemShut {NoStop}%
\bibitem [{\citenamefont {Chan}\ \emph {et~al.}(2011)\citenamefont {Chan},
  \citenamefont {Alegre}, \citenamefont {Safavi-Naeini}, \citenamefont {Hill},
  \citenamefont {Krause}, \citenamefont {Gr{\"o}blacher}, \citenamefont
  {Aspelmeyer},\ and\ \citenamefont {Painter}}]{Chan2011}%
  \BibitemOpen
  \bibfield  {author} {\bibinfo {author} {\bibfnamefont {J.}~\bibnamefont
  {Chan}}, \bibinfo {author} {\bibfnamefont {T.~P.~M.}\ \bibnamefont {Alegre}},
  \bibinfo {author} {\bibfnamefont {A.~H.}\ \bibnamefont {Safavi-Naeini}},
  \bibinfo {author} {\bibfnamefont {J.~T.}\ \bibnamefont {Hill}}, \bibinfo
  {author} {\bibfnamefont {A.}~\bibnamefont {Krause}}, \bibinfo {author}
  {\bibfnamefont {S.}~\bibnamefont {Gr{\"o}blacher}}, \bibinfo {author}
  {\bibfnamefont {M.}~\bibnamefont {Aspelmeyer}}, \ and\ \bibinfo {author}
  {\bibfnamefont {O.}~\bibnamefont {Painter}},\ }\bibfield  {title} {\enquote
  {\bibinfo {title} {{Laser cooling of a nanomechanical oscillator into its
  quantum ground state}},}\ }\href {\doibase 10.1038/nature10461} {\bibfield
  {journal} {\bibinfo  {journal} {Nature}\ }\textbf {\bibinfo {volume} {478}},\
  \bibinfo {pages} {89--92} (\bibinfo {year} {2011})}\BibitemShut {NoStop}%
\bibitem [{\citenamefont {Rimberg}\ \emph {et~al.}(2014)\citenamefont
  {Rimberg}, \citenamefont {Blencowe}, \citenamefont {Armour},\ and\
  \citenamefont {Nation}}]{Rimberg2014}%
  \BibitemOpen
  \bibfield  {author} {\bibinfo {author} {\bibfnamefont {A.~J.}\ \bibnamefont
  {Rimberg}}, \bibinfo {author} {\bibfnamefont {M.~P.}\ \bibnamefont
  {Blencowe}}, \bibinfo {author} {\bibfnamefont {A.~D.}\ \bibnamefont
  {Armour}}, \ and\ \bibinfo {author} {\bibfnamefont {P.~D.}\ \bibnamefont
  {Nation}},\ }\bibfield  {title} {\enquote {\bibinfo {title} {{A
  cavity-{C}ooper pair transistor scheme for investigating quantum
  optomechanics in the ultra-strong coupling regime}},}\ }\href {\doibase
  10.1088/1367-2630/16/5/055008} {\bibfield  {journal} {\bibinfo  {journal}
  {New J. Phys.}\ }\textbf {\bibinfo {volume} {16}},\ \bibinfo {pages} {055008}
  (\bibinfo {year} {2014})}\BibitemShut {NoStop}%
\bibitem [{\citenamefont {Aspelmeyer}\ \emph {et~al.}(2014)\citenamefont
  {Aspelmeyer}, \citenamefont {Kippenberg},\ and\ \citenamefont
  {Marquardt}}]{Aspelmeyer2014}%
  \BibitemOpen
  \bibfield  {author} {\bibinfo {author} {\bibfnamefont {M.}~\bibnamefont
  {Aspelmeyer}}, \bibinfo {author} {\bibfnamefont {T.~J.}\ \bibnamefont
  {Kippenberg}}, \ and\ \bibinfo {author} {\bibfnamefont {F.}~\bibnamefont
  {Marquardt}},\ }\bibfield  {title} {\enquote {\bibinfo {title} {{Cavity
  optomechanics}},}\ }\href {\doibase 10.1103/RevModPhys.86.1391} {\bibfield
  {journal} {\bibinfo  {journal} {Rev. Mod. Phys.}\ }\textbf {\bibinfo {volume}
  {86}},\ \bibinfo {pages} {1391--1452} (\bibinfo {year} {2014})}\BibitemShut
  {NoStop}%
\bibitem [{\citenamefont {Ridolfo}\ \emph {et~al.}(2012)\citenamefont
  {Ridolfo}, \citenamefont {Leib}, \citenamefont {Savasta},\ and\ \citenamefont
  {Hartmann}}]{Ridolfo2012}%
  \BibitemOpen
  \bibfield  {author} {\bibinfo {author} {\bibfnamefont {A.}~\bibnamefont
  {Ridolfo}}, \bibinfo {author} {\bibfnamefont {M.}~\bibnamefont {Leib}},
  \bibinfo {author} {\bibfnamefont {S.}~\bibnamefont {Savasta}}, \ and\
  \bibinfo {author} {\bibfnamefont {M.~J.}\ \bibnamefont {Hartmann}},\
  }\bibfield  {title} {\enquote {\bibinfo {title} {{Photon blockade in the
  ultrastrong coupling regime}},}\ }\href {\doibase
  10.1103/PhysRevLett.109.193602} {\bibfield  {journal} {\bibinfo  {journal}
  {Phys. Rev. Lett.}\ }\textbf {\bibinfo {volume} {109}},\ \bibinfo {pages}
  {193602} (\bibinfo {year} {2012})}\BibitemShut {NoStop}%
\bibitem [{\citenamefont {Garziano}\ \emph
  {et~al.}(2015{\natexlab{a}})\citenamefont {Garziano}, \citenamefont {Stassi},
  \citenamefont {Macr{\`{\i}}}, \citenamefont {Kockum}, \citenamefont
  {Savasta},\ and\ \citenamefont {Nori}}]{Garziano2015}%
  \BibitemOpen
  \bibfield  {author} {\bibinfo {author} {\bibfnamefont {L.}~\bibnamefont
  {Garziano}}, \bibinfo {author} {\bibfnamefont {R.}~\bibnamefont {Stassi}},
  \bibinfo {author} {\bibfnamefont {V.}~\bibnamefont {Macr{\`{\i}}}}, \bibinfo
  {author} {\bibfnamefont {A.~F.}\ \bibnamefont {Kockum}}, \bibinfo {author}
  {\bibfnamefont {S.}~\bibnamefont {Savasta}}, \ and\ \bibinfo {author}
  {\bibfnamefont {F.}~\bibnamefont {Nori}},\ }\bibfield  {title} {\enquote
  {\bibinfo {title} {{Multiphoton quantum Rabi oscillations in ultrastrong
  cavity QED}},}\ }\href {\doibase 10.1103/PhysRevA.92.063830} {\bibfield
  {journal} {\bibinfo  {journal} {Phys. Rev. A}\ }\textbf {\bibinfo {volume}
  {92}},\ \bibinfo {pages} {063830} (\bibinfo {year}
  {2015}{\natexlab{a}})}\BibitemShut {NoStop}%
\bibitem [{\citenamefont {Garziano}\ \emph {et~al.}(2016)\citenamefont
  {Garziano}, \citenamefont {Stassi}, \citenamefont {Macr{\`{\i}}},
  \citenamefont {{Di Stefano}}, \citenamefont {Nori},\ and\ \citenamefont
  {Savasta}}]{Garziano2016}%
  \BibitemOpen
  \bibfield  {author} {\bibinfo {author} {\bibfnamefont {L.}~\bibnamefont
  {Garziano}}, \bibinfo {author} {\bibfnamefont {R.}~\bibnamefont {Stassi}},
  \bibinfo {author} {\bibfnamefont {V.}~\bibnamefont {Macr{\`{\i}}}}, \bibinfo
  {author} {\bibfnamefont {O.}~\bibnamefont {{Di Stefano}}}, \bibinfo {author}
  {\bibfnamefont {F.}~\bibnamefont {Nori}}, \ and\ \bibinfo {author}
  {\bibfnamefont {S.}~\bibnamefont {Savasta}},\ }\bibfield  {title} {\enquote
  {\bibinfo {title} {{One photon can simultaneously excite two or more
  atoms}},}\ }\href {\doibase 10.1103/PhysRevLett.117.043601} {\bibfield
  {journal} {\bibinfo  {journal} {Phys. Rev. Lett.}\ }\textbf {\bibinfo
  {volume} {117}},\ \bibinfo {pages} {043601} (\bibinfo {year}
  {2016})}\BibitemShut {NoStop}%
\bibitem [{\citenamefont {Garziano}\ \emph
  {et~al.}(2015{\natexlab{b}})\citenamefont {Garziano}, \citenamefont {Stassi},
  \citenamefont {Macr\'{\i}}, \citenamefont {Savasta},\ and\ \citenamefont
  {Di~Stefano}}]{Garziano2015b}%
  \BibitemOpen
  \bibfield  {author} {\bibinfo {author} {\bibfnamefont {L.}~\bibnamefont
  {Garziano}}, \bibinfo {author} {\bibfnamefont {R.}~\bibnamefont {Stassi}},
  \bibinfo {author} {\bibfnamefont {V.}~\bibnamefont {Macr\'{\i}}}, \bibinfo
  {author} {\bibfnamefont {S.}~\bibnamefont {Savasta}}, \ and\ \bibinfo
  {author} {\bibfnamefont {O.}~\bibnamefont {Di~Stefano}},\ }\bibfield  {title}
  {\enquote {\bibinfo {title} {{Single-step arbitrary control of mechanical
  quantum states in ultrastrong optomechanics}},}\ }\href {\doibase
  10.1103/PhysRevA.91.023809} {\bibfield  {journal} {\bibinfo  {journal} {Phys.
  Rev. A}\ }\textbf {\bibinfo {volume} {91}},\ \bibinfo {pages} {023809}
  (\bibinfo {year} {2015}{\natexlab{b}})}\BibitemShut {NoStop}%
\bibitem [{\citenamefont {Macr\'{\i}}\ \emph {et~al.}(2016)\citenamefont
  {Macr\'{\i}}, \citenamefont {Garziano}, \citenamefont {Ridolfo},
  \citenamefont {Di~Stefano},\ and\ \citenamefont {Savasta}}]{Macri2016}%
  \BibitemOpen
  \bibfield  {author} {\bibinfo {author} {\bibfnamefont {V.}~\bibnamefont
  {Macr\'{\i}}}, \bibinfo {author} {\bibfnamefont {L.}~\bibnamefont
  {Garziano}}, \bibinfo {author} {\bibfnamefont {A.}~\bibnamefont {Ridolfo}},
  \bibinfo {author} {\bibfnamefont {O.}~\bibnamefont {Di~Stefano}}, \ and\
  \bibinfo {author} {\bibfnamefont {S.}~\bibnamefont {Savasta}},\ }\bibfield
  {title} {\enquote {\bibinfo {title} {{Deterministic synthesis of mechanical
  {NOON} states in ultrastrong optomechanics}},}\ }\href {\doibase
  10.1103/PhysRevA.94.013817} {\bibfield  {journal} {\bibinfo  {journal} {Phys.
  Rev. A}\ }\textbf {\bibinfo {volume} {94}},\ \bibinfo {pages} {013817}
  (\bibinfo {year} {2016})}\BibitemShut {NoStop}%
\bibitem [{\citenamefont {Butera}\ and\ \citenamefont
  {Carusotto}(2018)}]{Butera2018}%
  \BibitemOpen
  \bibfield  {author} {\bibinfo {author} {\bibfnamefont {S.}~\bibnamefont
  {Butera}}\ and\ \bibinfo {author} {\bibfnamefont {I.}~\bibnamefont
  {Carusotto}},\ }\bibfield  {title} {\enquote {\bibinfo {title} {{Mechanical
  back-reaction effect of the dynamical Casimir emission}},}\ }\href
  {http://arxiv.org/abs/1810.11281} {\  (\bibinfo {year} {2018})},\ \Eprint
  {http://arxiv.org/abs/1810.11281} {arXiv:1810.11281} \BibitemShut {NoStop}%
\bibitem [{\citenamefont {Plunien}\ \emph {et~al.}(1987)\citenamefont
  {Plunien}, \citenamefont {M\"uller},\ and\ \citenamefont
  {Greiner}}]{PLUNIEN1987}%
  \BibitemOpen
  \bibfield  {author} {\bibinfo {author} {\bibfnamefont {G.}~\bibnamefont
  {Plunien}}, \bibinfo {author} {\bibfnamefont {B.}~\bibnamefont {M\"uller}}, \
  and\ \bibinfo {author} {\bibfnamefont {W.}~\bibnamefont {Greiner}},\
  }\bibfield  {title} {\enquote {\bibinfo {title} {{Casimir energy at finite
  temperature}},}\ }\href
  {http://www.sciencedirect.com/science/article/pii/0378437187902470}
  {\bibfield  {journal} {\bibinfo  {journal} {Physica A}\ }\textbf {\bibinfo
  {volume} {145}},\ \bibinfo {pages} {202 -- 219} (\bibinfo {year}
  {1987})}\BibitemShut {NoStop}%
\bibitem [{\citenamefont {Sch\"utzhold}\ \emph {et~al.}(1998)\citenamefont
  {Sch\"utzhold}, \citenamefont {Plunien},\ and\ \citenamefont
  {Soff}}]{Plunien1998}%
  \BibitemOpen
  \bibfield  {author} {\bibinfo {author} {\bibfnamefont {R.}~\bibnamefont
  {Sch\"utzhold}}, \bibinfo {author} {\bibfnamefont {G.}~\bibnamefont
  {Plunien}}, \ and\ \bibinfo {author} {\bibfnamefont {G.}~\bibnamefont
  {Soff}},\ }\bibfield  {title} {\enquote {\bibinfo {title} {{Quantum radiation
  in external background fields}},}\ }\href {\doibase 10.1103/PhysRevA.58.1783}
  {\bibfield  {journal} {\bibinfo  {journal} {Phys. Rev. A}\ }\textbf {\bibinfo
  {volume} {58}},\ \bibinfo {pages} {1783--1793} (\bibinfo {year}
  {1998})}\BibitemShut {NoStop}%
\bibitem [{\citenamefont {Plunien}\ \emph {et~al.}(2000)\citenamefont
  {Plunien}, \citenamefont {Sch\"utzhold},\ and\ \citenamefont
  {Soff}}]{Plunien2000}%
  \BibitemOpen
  \bibfield  {author} {\bibinfo {author} {\bibfnamefont {G.}~\bibnamefont
  {Plunien}}, \bibinfo {author} {\bibfnamefont {R.}~\bibnamefont
  {Sch\"utzhold}}, \ and\ \bibinfo {author} {\bibfnamefont {G.}~\bibnamefont
  {Soff}},\ }\bibfield  {title} {\enquote {\bibinfo {title} {{Dynamical Casimir
  effect at finite temperature}},}\ }\href {\doibase
  10.1103/PhysRevLett.84.1882} {\bibfield  {journal} {\bibinfo  {journal}
  {Phys. Rev. Lett.}\ }\textbf {\bibinfo {volume} {84}},\ \bibinfo {pages}
  {1882--1885} (\bibinfo {year} {2000})}\BibitemShut {NoStop}%
\bibitem [{\citenamefont {Motazedifard}\ \emph {et~al.}(2018)\citenamefont
  {Motazedifard}, \citenamefont {Dalafi}, \citenamefont {Naderi},\ and\
  \citenamefont {Roknizadeh}}]{MOTAZEDIFARD2018}%
  \BibitemOpen
  \bibfield  {author} {\bibinfo {author} {\bibfnamefont {A.}~\bibnamefont
  {Motazedifard}}, \bibinfo {author} {\bibfnamefont {A.}~\bibnamefont
  {Dalafi}}, \bibinfo {author} {\bibfnamefont {M.H.}\ \bibnamefont {Naderi}}, \
  and\ \bibinfo {author} {\bibfnamefont {R.}~\bibnamefont {Roknizadeh}},\
  }\bibfield  {title} {\enquote {\bibinfo {title} {{Controllable generation of
  photons and phonons in a coupled Bose-Einstein condensate-optomechanical
  cavity via the parametric dynamical Casimir effect}},}\ }\href
  {http://www.sciencedirect.com/science/article/pii/S0003491618301842}
  {\bibfield  {journal} {\bibinfo  {journal} {Ann. Physics}\ }\textbf {\bibinfo
  {volume} {396}},\ \bibinfo {pages} {202 -- 219} (\bibinfo {year}
  {2018})}\BibitemShut {NoStop}%
\bibitem [{\citenamefont {{Di Stefano}}\ \emph {et~al.}(2017)\citenamefont {{Di
  Stefano}}, \citenamefont {Stassi}, \citenamefont {Garziano}, \citenamefont
  {Kockum}, \citenamefont {Savasta},\ and\ \citenamefont
  {Nori}}]{DiStefano2017}%
  \BibitemOpen
  \bibfield  {author} {\bibinfo {author} {\bibfnamefont {O.}~\bibnamefont {{Di
  Stefano}}}, \bibinfo {author} {\bibfnamefont {R.}~\bibnamefont {Stassi}},
  \bibinfo {author} {\bibfnamefont {L.}~\bibnamefont {Garziano}}, \bibinfo
  {author} {\bibfnamefont {A.~F.}\ \bibnamefont {Kockum}}, \bibinfo {author}
  {\bibfnamefont {S.}~\bibnamefont {Savasta}}, \ and\ \bibinfo {author}
  {\bibfnamefont {F.}~\bibnamefont {Nori}},\ }\bibfield  {title} {\enquote
  {\bibinfo {title} {{Feynman-diagrams approach to the quantum Rabi model for
  ultrastrong cavity QED: stimulated emission and reabsorption of virtual
  particles dressing a physical excitation}},}\ }\href
  {http://iopscience.iop.org/article/10.1088/1367-2630/aa6cd7/meta#} {\bibfield
   {journal} {\bibinfo  {journal} {New J. Phys.}\ }\textbf {\bibinfo {volume}
  {19}},\ \bibinfo {pages} {053010} (\bibinfo {year} {2017})}\BibitemShut
  {NoStop}%
\bibitem [{\citenamefont {Stassi}\ \emph {et~al.}(2017)\citenamefont {Stassi},
  \citenamefont {Macr\`{\i}}, \citenamefont {Kockum}, \citenamefont
  {Di~Stefano}, \citenamefont {Miranowicz}, \citenamefont {Savasta},\ and\
  \citenamefont {Nori}}]{Stassi2017}%
  \BibitemOpen
  \bibfield  {author} {\bibinfo {author} {\bibfnamefont {R.}~\bibnamefont
  {Stassi}}, \bibinfo {author} {\bibfnamefont {V.}~\bibnamefont {Macr\`{\i}}},
  \bibinfo {author} {\bibfnamefont {A.~F.}\ \bibnamefont {Kockum}}, \bibinfo
  {author} {\bibfnamefont {O.}~\bibnamefont {Di~Stefano}}, \bibinfo {author}
  {\bibfnamefont {A.}~\bibnamefont {Miranowicz}}, \bibinfo {author}
  {\bibfnamefont {S.}~\bibnamefont {Savasta}}, \ and\ \bibinfo {author}
  {\bibfnamefont {F.}~\bibnamefont {Nori}},\ }\bibfield  {title} {\enquote
  {\bibinfo {title} {{Quantum nonlinear optics without photons}},}\ }\href
  {\doibase 10.1103/PhysRevA.96.023818} {\bibfield  {journal} {\bibinfo
  {journal} {Phys. Rev. A}\ }\textbf {\bibinfo {volume} {96}},\ \bibinfo
  {pages} {023818} (\bibinfo {year} {2017})}\BibitemShut {NoStop}%
\bibitem [{\citenamefont {Kockum}\ \emph {et~al.}(2017)\citenamefont {Kockum},
  \citenamefont {Miranowicz}, \citenamefont {Macr\`{\i}}, \citenamefont
  {Savasta},\ and\ \citenamefont {Nori}}]{Kockum2017}%
  \BibitemOpen
  \bibfield  {author} {\bibinfo {author} {\bibfnamefont {A.~F.}\ \bibnamefont
  {Kockum}}, \bibinfo {author} {\bibfnamefont {A.}~\bibnamefont {Miranowicz}},
  \bibinfo {author} {\bibfnamefont {V.}~\bibnamefont {Macr\`{\i}}}, \bibinfo
  {author} {\bibfnamefont {S.}~\bibnamefont {Savasta}}, \ and\ \bibinfo
  {author} {\bibfnamefont {F.}~\bibnamefont {Nori}},\ }\bibfield  {title}
  {\enquote {\bibinfo {title} {{Deterministic quantum nonlinear optics with
  single atoms and virtual photons}},}\ }\href {\doibase
  10.1103/PhysRevA.95.063849} {\bibfield  {journal} {\bibinfo  {journal} {Phys.
  Rev. A}\ }\textbf {\bibinfo {volume} {95}},\ \bibinfo {pages} {063849}
  (\bibinfo {year} {2017})}\BibitemShut {NoStop}%
\bibitem [{\citenamefont {Kockum}\ \emph {et~al.}(2019)\citenamefont {Kockum},
  \citenamefont {Miranowicz}, \citenamefont {De~Liberato}, \citenamefont
  {Savasta},\ and\ \citenamefont {Nori}}]{Kockum2018}%
  \BibitemOpen
  \bibfield  {author} {\bibinfo {author} {\bibfnamefont {A.~F.}\ \bibnamefont
  {Kockum}}, \bibinfo {author} {\bibfnamefont {A.}~\bibnamefont {Miranowicz}},
  \bibinfo {author} {\bibfnamefont {S.}~\bibnamefont {De~Liberato}}, \bibinfo
  {author} {\bibfnamefont {S.}~\bibnamefont {Savasta}}, \ and\ \bibinfo
  {author} {\bibfnamefont {F.}~\bibnamefont {Nori}},\ }\bibfield  {title}
  {\enquote {\bibinfo {title} {{Ultrastrong coupling between light and
  matter}},}\ }\href {https://www.nature.com/articles/s42254-018-0006-2}
  {\bibfield  {journal} {\bibinfo  {journal} {Nat. Rev. Phys.}\ }\textbf
  {\bibinfo {volume} {1}},\ \bibinfo {pages} {19--40} (\bibinfo {year}
  {2019})}\BibitemShut {NoStop}%
\bibitem [{\citenamefont {Forn-D{\'{i}}az}\ \emph {et~al.}(2018)\citenamefont
  {Forn-D{\'{i}}az}, \citenamefont {Lamata}, \citenamefont {Rico},
  \citenamefont {Kono},\ and\ \citenamefont {Solano}}]{Forn-Diaz2018a}%
  \BibitemOpen
  \bibfield  {author} {\bibinfo {author} {\bibfnamefont {P.}~\bibnamefont
  {Forn-D{\'{i}}az}}, \bibinfo {author} {\bibfnamefont {L.}~\bibnamefont
  {Lamata}}, \bibinfo {author} {\bibfnamefont {E.}~\bibnamefont {Rico}},
  \bibinfo {author} {\bibfnamefont {J.}~\bibnamefont {Kono}}, \ and\ \bibinfo
  {author} {\bibfnamefont {E.}~\bibnamefont {Solano}},\ }\bibfield  {title}
  {\enquote {\bibinfo {title} {{Ultrastrong coupling regimes of light-matter
  interaction}},}\ }\href {http://arxiv.org/abs/1804.09275} {\  (\bibinfo
  {year} {2018})},\ \Eprint {http://arxiv.org/abs/1804.09275}
  {arXiv:1804.09275} \BibitemShut {NoStop}%
\bibitem [{\citenamefont {Beaudoin}\ \emph {et~al.}(2011)\citenamefont
  {Beaudoin}, \citenamefont {Gambetta},\ and\ \citenamefont
  {Blais}}]{Beaudoin2011}%
  \BibitemOpen
  \bibfield  {author} {\bibinfo {author} {\bibfnamefont {F.}~\bibnamefont
  {Beaudoin}}, \bibinfo {author} {\bibfnamefont {J.~M.}\ \bibnamefont
  {Gambetta}}, \ and\ \bibinfo {author} {\bibfnamefont {A.}~\bibnamefont
  {Blais}},\ }\bibfield  {title} {\enquote {\bibinfo {title} {{Dissipation and
  ultrastrong coupling in circuit QED}},}\ }\href {\doibase
  10.1103/PhysRevA.84.043832} {\bibfield  {journal} {\bibinfo  {journal} {Phys.
  Rev. A}\ }\textbf {\bibinfo {volume} {84}},\ \bibinfo {pages} {043832}
  (\bibinfo {year} {2011})}\BibitemShut {NoStop}%
\bibitem [{\citenamefont {Hu}\ \emph {et~al.}(2015)\citenamefont {Hu},
  \citenamefont {Huang}, \citenamefont {Liao}, \citenamefont {Tian},\ and\
  \citenamefont {Goan}}]{Hu2015}%
  \BibitemOpen
  \bibfield  {author} {\bibinfo {author} {\bibfnamefont {D.}~\bibnamefont
  {Hu}}, \bibinfo {author} {\bibfnamefont {S.-Y.}\ \bibnamefont {Huang}},
  \bibinfo {author} {\bibfnamefont {J.-Q.}\ \bibnamefont {Liao}}, \bibinfo
  {author} {\bibfnamefont {L.}~\bibnamefont {Tian}}, \ and\ \bibinfo {author}
  {\bibfnamefont {H.-S.}\ \bibnamefont {Goan}},\ }\bibfield  {title} {\enquote
  {\bibinfo {title} {{Quantum coherence in ultrastrong optomechanics}},}\
  }\href {\doibase 10.1103/PhysRevA.91.013812} {\bibfield  {journal} {\bibinfo
  {journal} {Phys. Rev. A}\ }\textbf {\bibinfo {volume} {91}},\ \bibinfo
  {pages} {013812} (\bibinfo {year} {2015})}\BibitemShut {NoStop}%
\bibitem [{\citenamefont {Ma}\ and\ \citenamefont {Law}(2015)}]{Ma2015}%
  \BibitemOpen
  \bibfield  {author} {\bibinfo {author} {\bibfnamefont {K.~W.}\ \bibnamefont
  {Ma}}\ and\ \bibinfo {author} {\bibfnamefont {C.~K.}\ \bibnamefont {Law}},\
  }\bibfield  {title} {\enquote {\bibinfo {title} {{Three-photon resonance and
  adiabatic passage in the large-detuning {R}abi model}},}\ }\href {\doibase
  10.1103/PhysRevA.92.023842} {\bibfield  {journal} {\bibinfo  {journal} {Phys.
  Rev. A}\ }\textbf {\bibinfo {volume} {92}},\ \bibinfo {pages} {023842}
  (\bibinfo {year} {2015})}\BibitemShut {NoStop}%
\bibitem [{\citenamefont {Settineri}\ \emph {et~al.}(2018)\citenamefont
  {Settineri}, \citenamefont {Macr\'{\i}}, \citenamefont {Ridolfo},
  \citenamefont {{Di Stefano}}, \citenamefont {Kockum}, \citenamefont {Nori},\
  and\ \citenamefont {Savasta}}]{settineri2018}%
  \BibitemOpen
  \bibfield  {author} {\bibinfo {author} {\bibfnamefont {A.}~\bibnamefont
  {Settineri}}, \bibinfo {author} {\bibfnamefont {V.}~\bibnamefont
  {Macr\'{\i}}}, \bibinfo {author} {\bibfnamefont {A.}~\bibnamefont {Ridolfo}},
  \bibinfo {author} {\bibfnamefont {O.}~\bibnamefont {{Di Stefano}}}, \bibinfo
  {author} {\bibfnamefont {A.~F.}\ \bibnamefont {Kockum}}, \bibinfo {author}
  {\bibfnamefont {F.}~\bibnamefont {Nori}}, \ and\ \bibinfo {author}
  {\bibfnamefont {S.}~\bibnamefont {Savasta}},\ }\bibfield  {title} {\enquote
  {\bibinfo {title} {{Dissipation and thermal noise in hybrid quantum systems
  in the ultrastrong-coupling regime}},}\ }\href {\doibase
  10.1103/PhysRevA.98.053834} {\bibfield  {journal} {\bibinfo  {journal} {Phys.
  Rev. A}\ }\textbf {\bibinfo {volume} {98}},\ \bibinfo {pages} {053834}
  (\bibinfo {year} {2018})}\BibitemShut {NoStop}%
\bibitem [{\citenamefont {Portolan}\ \emph {et~al.}(2008)\citenamefont
  {Portolan}, \citenamefont {Di~Stefano}, \citenamefont {Savasta},
  \citenamefont {Rossi},\ and\ \citenamefont {Girlanda}}]{Portolan2008}%
  \BibitemOpen
  \bibfield  {author} {\bibinfo {author} {\bibfnamefont {S.}~\bibnamefont
  {Portolan}}, \bibinfo {author} {\bibfnamefont {O.}~\bibnamefont
  {Di~Stefano}}, \bibinfo {author} {\bibfnamefont {S.}~\bibnamefont {Savasta}},
  \bibinfo {author} {\bibfnamefont {F.}~\bibnamefont {Rossi}}, \ and\ \bibinfo
  {author} {\bibfnamefont {R.}~\bibnamefont {Girlanda}},\ }\bibfield  {title}
  {\enquote {\bibinfo {title} {{Nonequilibrium Langevin approach to quantum
  optics in semiconductor microcavities}},}\ }\href
  {https://link.aps.org/doi/10.1103/PhysRevB.77.035433} {\bibfield  {journal}
  {\bibinfo  {journal} {Phys. Rev. B}\ }\textbf {\bibinfo {volume} {77}},\
  \bibinfo {pages} {035433} (\bibinfo {year} {2008})}\BibitemShut {NoStop}%
\bibitem [{\citenamefont {Di~Stefano}\ \emph {et~al.}(2001)\citenamefont
  {Di~Stefano}, \citenamefont {Savasta},\ and\ \citenamefont
  {Girlanda}}]{stefano2001}%
  \BibitemOpen
  \bibfield  {author} {\bibinfo {author} {\bibfnamefont {O.}~\bibnamefont
  {Di~Stefano}}, \bibinfo {author} {\bibfnamefont {S.}~\bibnamefont {Savasta}},
  \ and\ \bibinfo {author} {\bibfnamefont {R.}~\bibnamefont {Girlanda}},\
  }\bibfield  {title} {\enquote {\bibinfo {title} {{Mode expansion and photon
  operators in dispersive and absorbing dielectrics}},}\ }\href
  {https://www.tandfonline.com/doi/abs/10.1080/09500340108235155} {\bibfield
  {journal} {\bibinfo  {journal} {J. Mod. Opt.}\ }\textbf {\bibinfo {volume}
  {48}},\ \bibinfo {pages} {67--84} (\bibinfo {year} {2001})}\BibitemShut
  {NoStop}%
\bibitem [{\citenamefont {Gardiner}\ and\ \citenamefont
  {Collett}(1985)}]{Gardiner1985}%
  \BibitemOpen
  \bibfield  {author} {\bibinfo {author} {\bibfnamefont {C.~W.}\ \bibnamefont
  {Gardiner}}\ and\ \bibinfo {author} {\bibfnamefont {M.~J.}\ \bibnamefont
  {Collett}},\ }\bibfield  {title} {\enquote {\bibinfo {title} {{Input and
  output in damped quantum systems: quantum stochastic differential equations
  and the master equation}},}\ }\href {\doibase 10.1103/PhysRevA.31.3761}
  {\bibfield  {journal} {\bibinfo  {journal} {Phys. Rev. A}\ }\textbf {\bibinfo
  {volume} {31}},\ \bibinfo {pages} {3761--3774} (\bibinfo {year}
  {1985})}\BibitemShut {NoStop}%
\bibitem [{\citenamefont {Savasta}\ \emph {et~al.}(2002)\citenamefont
  {Savasta}, \citenamefont {{Di Stefano}},\ and\ \citenamefont
  {Girlanda}}]{Savasta2002}%
  \BibitemOpen
  \bibfield  {author} {\bibinfo {author} {\bibfnamefont {S.}~\bibnamefont
  {Savasta}}, \bibinfo {author} {\bibfnamefont {O.}~\bibnamefont {{Di
  Stefano}}}, \ and\ \bibinfo {author} {\bibfnamefont {R.}~\bibnamefont
  {Girlanda}},\ }\bibfield  {title} {\enquote {\bibinfo {title} {{Light
  quantization for arbitrary scattering systems}},}\ }\href {\doibase
  10.1103/PhysRevA.65.043801} {\bibfield  {journal} {\bibinfo  {journal} {Phys.
  Rev. A}\ }\textbf {\bibinfo {volume} {65}},\ \bibinfo {pages} {043801}
  (\bibinfo {year} {2002})}\BibitemShut {NoStop}%
\bibitem [{\citenamefont {{Di Stefano}}\ \emph {et~al.}(2018)\citenamefont {{Di
  Stefano}}, \citenamefont {Kockum}, \citenamefont {Ridolfo}, \citenamefont
  {Savasta},\ and\ \citenamefont {Nori}}]{DiStefano2018}%
  \BibitemOpen
  \bibfield  {author} {\bibinfo {author} {\bibfnamefont {O.}~\bibnamefont {{Di
  Stefano}}}, \bibinfo {author} {\bibfnamefont {A.~F.}\ \bibnamefont {Kockum}},
  \bibinfo {author} {\bibfnamefont {A.}~\bibnamefont {Ridolfo}}, \bibinfo
  {author} {\bibfnamefont {S.}~\bibnamefont {Savasta}}, \ and\ \bibinfo
  {author} {\bibfnamefont {F.}~\bibnamefont {Nori}},\ }\bibfield  {title}
  {\enquote {\bibinfo {title} {{Photodetection probability in quantum systems
  with arbitrarily strong light-matter interaction}},}\ }\href
  {https://www.nature.com/articles/s41598-018-36056-1} {\bibfield  {journal}
  {\bibinfo  {journal} {Sci. Rep.}\ }\textbf {\bibinfo {volume} {8}},\ \bibinfo
  {pages} {17825} (\bibinfo {year} {2018})}\BibitemShut {NoStop}%
\bibitem [{\citenamefont {O'Connell}\ \emph {et~al.}(2010)\citenamefont
  {O'Connell}, \citenamefont {Hofheinz}, \citenamefont {Ansmann}, \citenamefont
  {Bialczak}, \citenamefont {Lenander}, \citenamefont {Lucero}, \citenamefont
  {Neeley}, \citenamefont {Sank}, \citenamefont {Wang}, \citenamefont {Weides},
  \citenamefont {Wenner}, \citenamefont {Martinis},\ and\ \citenamefont
  {Cleland}}]{OConnell2010}%
  \BibitemOpen
  \bibfield  {author} {\bibinfo {author} {\bibfnamefont {A.D.}\ \bibnamefont
  {O'Connell}}, \bibinfo {author} {\bibfnamefont {M.}~\bibnamefont {Hofheinz}},
  \bibinfo {author} {\bibfnamefont {M.}~\bibnamefont {Ansmann}}, \bibinfo
  {author} {\bibfnamefont {R.C.}\ \bibnamefont {Bialczak}}, \bibinfo {author}
  {\bibfnamefont {M.}~\bibnamefont {Lenander}}, \bibinfo {author}
  {\bibfnamefont {E.}~\bibnamefont {Lucero}}, \bibinfo {author} {\bibfnamefont
  {M.}~\bibnamefont {Neeley}}, \bibinfo {author} {\bibfnamefont
  {D.}~\bibnamefont {Sank}}, \bibinfo {author} {\bibfnamefont {H.}~\bibnamefont
  {Wang}}, \bibinfo {author} {\bibfnamefont {M.}~\bibnamefont {Weides}},
  \bibinfo {author} {\bibfnamefont {J.}~\bibnamefont {Wenner}}, \bibinfo
  {author} {\bibfnamefont {J.M.}\ \bibnamefont {Martinis}}, \ and\ \bibinfo
  {author} {\bibfnamefont {A.N.}\ \bibnamefont {Cleland}},\ }\bibfield  {title}
  {\enquote {\bibinfo {title} {{Quantum ground state and single-phonon control
  of a mechanical resonator}},}\ }\href {\doibase 10.1038/nature08967}
  {\bibfield  {journal} {\bibinfo  {journal} {Nature}\ }\textbf {\bibinfo
  {volume} {464}},\ \bibinfo {pages} {697--703} (\bibinfo {year}
  {2010})}\BibitemShut {NoStop}%
\bibitem [{\citenamefont {Rouxinol}\ \emph {et~al.}(2016)\citenamefont
  {Rouxinol}, \citenamefont {Hao}, \citenamefont {Brito}, \citenamefont
  {Caldeira}, \citenamefont {Irish},\ and\ \citenamefont {{La
  Haye}}}]{Rouxinol2016}%
  \BibitemOpen
  \bibfield  {author} {\bibinfo {author} {\bibfnamefont {F.}~\bibnamefont
  {Rouxinol}}, \bibinfo {author} {\bibfnamefont {Y.}~\bibnamefont {Hao}},
  \bibinfo {author} {\bibfnamefont {F.}~\bibnamefont {Brito}}, \bibinfo
  {author} {\bibfnamefont {A.~O.}\ \bibnamefont {Caldeira}}, \bibinfo {author}
  {\bibfnamefont {E.~K.}\ \bibnamefont {Irish}}, \ and\ \bibinfo {author}
  {\bibfnamefont {M.~D.}\ \bibnamefont {{La Haye}}},\ }\bibfield  {title}
  {\enquote {\bibinfo {title} {{Measurements of nanoresonator-qubit
  interactions in a hybrid quantum electromechanical system}},}\ }\href
  {\doibase 10.1088/0957-4484/27/36/364003} {\bibfield  {journal} {\bibinfo
  {journal} {Nanotechnology}\ }\textbf {\bibinfo {volume} {27}},\ \bibinfo
  {pages} {364003} (\bibinfo {year} {2016})}\BibitemShut {NoStop}%
\bibitem [{\citenamefont {Qin}\ \emph {et~al.}(2019)\citenamefont {Qin},
  \citenamefont {Macr{\`\i}}, \citenamefont {Miranowicz}, \citenamefont
  {Savasta},\ and\ \citenamefont {Nori}}]{qin2019}%
  \BibitemOpen
  \bibfield  {author} {\bibinfo {author} {\bibfnamefont {W.}~\bibnamefont
  {Qin}}, \bibinfo {author} {\bibfnamefont {V.}~\bibnamefont {Macr{\`\i}}},
  \bibinfo {author} {\bibfnamefont {A.}~\bibnamefont {Miranowicz}}, \bibinfo
  {author} {\bibfnamefont {S.}~\bibnamefont {Savasta}}, \ and\ \bibinfo
  {author} {\bibfnamefont {F.}~\bibnamefont {Nori}},\ }\bibfield  {title}
  {\enquote {\bibinfo {title} {{Experimentally Feasible Dynamical Casimir
  Effect in Parametrically Amplified Cavity Optomechanics}},}\ }\href
  {https://arxiv.org/abs/1902.04216} {\bibfield  {journal} {\bibinfo  {journal}
  {arXiv:1902.04216}\ } (\bibinfo {year} {2019})}\BibitemShut {NoStop}%
\end{thebibliography}%
	
\end{document}